\documentclass[12pt, twoside]{article}
\pdfoutput=1

\newif\ifispreprint
\ispreprinttrue % <--------------------------- PREPRINT TOGGLE
\newcommand{\ifpreprintelse}[2]{\ifispreprint #1 \else #2 \fi}

\makeatletter
\newcommand{\affiliation}[1]{\gdef\@affiliation{#1}}
\renewcommand\maketitle{
  \begin{center}
    {\Large \textbf{\@title} \par} % Title font size
    \vskip 1em
    {\normalsize \@author \par} % Author font size
    \vskip 0.5em
    {\small \@affiliation \par} % Affiliation font size
    \vskip 1em
  \end{center}
}
\makeatother

\title{Partitioning of multiple brain metastases improves dose gradients in single-isocenter radiosurgery}
\author{
Johan~Sundström\textsuperscript{1},
Anton~Finnson\textsuperscript{1}, 
Elin~Hynning\textsuperscript{1},
Geert~De~Kerf\textsuperscript{2,3}, \\
Albin~Fredriksson\textsuperscript{1}
}
\affiliation{
\textsuperscript{1}RaySearch Laboratories, Stockholm, Sweden\\
\textsuperscript{2}Department of Radiation Oncology, Iridium Netwerk, Wilrijk (Antwerp), Belgium \\
\textsuperscript{3}Faculty of Medicine and Health Sciences, University of Antwerp, Antwerp, Belgium. 
}

\usepackage{fancyhdr}
\renewcommand{\thesubsection}{\thesection{\sf \Alph{subsection}}.}

\usepackage[toc,page]{appendix}
\newenvironment{customappendicesenvironment}[2]
{% begin code
    \appendixpageoff
    \appendixtitleoff
    \renewcommand{\appendixtocname}{{#1}}
    \begin{appendices}
    
    \renewcommand{\thesubsection}{{#2}}
    \numberwithin{table}{subsection}
    \numberwithin{figure}{subsection}
    \numberwithin{equation}{subsection}
    \noindent\ignorespaces
    \section*{#1}
}
{% end code
    \end{appendices}\ignorespacesafterend
}

\usepackage{graphicx}

\newcommand{\captionv}[3]
{
\begin{center}
\parbox{#1cm}
{
    \caption[#2]{{\sf #3}}
}
\end{center}
}

\usepackage[section]{placeins}
\usepackage[super,sort,comma]{natbib}
\makeatletter \renewcommand\@biblabel[1]{$^{#1}$} \makeatother
\usepackage{hyperref}
\hypersetup{colorlinks, citecolor=blue, filecolor=blue, linkcolor=blue, urlcolor=blue}

\usepackage[all]{hypcap}  % Causes link to figures to go to figure, not caption
\usepackage{amsmath}
\usepackage{amssymb}

\usepackage{siunitx} % \unit and table alignment
\usepackage{booktabs} % \toprule, \midrule, \bottomrule
\usepackage{multirow}

\usepackage[version=4]{mhchem} % cobolt 60
\usepackage{mathrsfs} % For \mathscr used in \setofpartitions

\newcommand{\floor}[1]{\left\lfloor #1 \right\rfloor}
\newcommand{\abs}[1]{\left\lvert #1 \right\rvert}
\newcommand{\setofpartitions}[2]{\mathscr{P}({#1}, {#2})}
\newcommand{\vtwelve}{\text{V}_{12~\unit{\gray}}}
\newcommand{\doseatpercentvolume}[1]{$\text{D}_{#1\%}$}
\newcommand{\doseatccvolume}[1]{$\text{D}_{#1~\unit{\centi\meter\cubed}}$}

\ifpreprintelse{
}{
\usepackage[mathlines]{lineno}
\linenumbers
}

\begin{document}

\maketitle

\pagenumbering{roman}
\setcounter{page}{1}
\pagestyle{plain}

\ifpreprintelse{}{
Corresponding author: Johan Sundström
\clearpage
}

\begin{abstract}
\noindent{\bf Background:}
A growing number of cancer patients with brain metastases can benefit from stereotactic radiosurgery (SRS) thanks to recent advances in systemic therapies which have led to improved survival.
Meanwhile, selection criteria for SRS treatments are evolving to include patients with increasingly many metastases.
With an increasing patient load, single-isocenter treatments on widely available C-arm linear accelerators are an attractive option. 
However, the planning of such treatments is challenging for multi-target cases due to the island blocking problem, which occurs when the multi-leaf collimator cannot conform to all targets simultaneously.
\\ 

\noindent {\bf Purpose:}
We propose a multi-target partitioning algorithm that mitigates excessive exposure of normal tissue caused by the island blocking problem.
\\

\noindent {\bf Methods:}
The proposed algorithm considers an initial set of arc trajectories and divides (partitions) the set of targets per trajectory into smaller subsets to treat with separate back-and-forth arc passes, simultaneously optimizing both the target subsets and collimator angles to minimize island blocking. We incorporated this algorithm into a fully automated treatment planning script and evaluated it on 20 simulated patient cases, each with 10 brain metastases and 21~\unit{\gray} prescriptions. For each case, the script generated a series of volumetric modulated arc therapy (VMAT) plans with increasingly many arcs along the three trajectories. Each such plan was compared to four baseline plans generated with alternative heuristics for distributing targets across arcs. We also evaluated the algorithm retrospectively on six clinical cases.
\\

\noindent {\bf Results:}
Partitioning significantly improved the gradient index (GI), global efficiency index (G$\eta$) and brain $\vtwelve$ compared to simultaneous treatment of all metastases. For example, the average GI improved from 5.9 to 3.3, G$\eta$ from 0.32 to 0.46, and normal brain $\vtwelve$ from 49~\unit{\centi\meter\cubed} to 26~\unit{\centi\meter\cubed} between 3 and 9 arcs. The baseline plans improved similarly, but the proposed algorithm was significantly better at utilizing a limited budget of arcs. All target partitioning strategies increased the total number of monitor units (MUs).
\\

\noindent {\bf Conclusions:}
The dose gradient in single-isocenter VMAT plans can be substantially improved by treating a smaller subset of metastases at a time. This requires more MUs and arcs, implying a trade-off between delivery time and plan quality which can be explored using the algorithm proposed in this paper.

\end{abstract}

\clearpage

% The table of contents is for drafting and refereeing purposes only.
% \tableofcontents
% \clearpage

\ifpreprintelse{
% No extra spacing in pre-print
}{
\setlength{\baselineskip}{0.7cm}  % double spacing
}
\pagenumbering{arabic}
\setcounter{page}{1}
\pagestyle{fancy}

\section{Introduction}
Brain metastases (BMs) develop in a large portion of adult cancer patients.
The exact incidence is unknown, but estimates vary between 10--40\unit{\percent}.\cite{lin2015treatment, lamba2021epidemiology} 
Historically, all patients with BMs have been given a very poor prognosis.\cite{lin2015treatment} 
However, over the last few decades, the heterogeneity of the patient group has been increasingly recognized,\cite{gaspar2000validation, sperduto2012summary}
which has led to increasingly individualized treatments. 
The median survival now varies from a couple of months to several years depending on various prognostic factors.\cite{lin2015treatment, suh2020current}
For patients with better prognoses, this has shifted the focus from short-term palliative treatments towards quality of life and long-term survival. \cite{lin2015treatment}
In radiotherapy, there has been a corresponding shift away from whole-brain radiotherapy (WBRT) towards stereotactic radiosurgery (SRS) since WBRT is associated with progressive neurocognitive decline.\cite{lin2015treatment, achrol2019brain, suh2020current}
% For mainly practical reasons, 
The application of SRS was initially limited to cases with a small number of BMs, but during the last few decades, technical advancements have enabled the treatment of larger numbers of metastases, and the clinical evidence for such treatments is evolving.\cite{achrol2019brain, suh2020current}
Guidelines already recommend SRS alone for up to four newly diagnosed BMs,\cite{suh2020current, vogelbaum2022treatment} a prospective study showed non-inferior results with SRS for 5--10 BMs compared to 2--4 BMs, \cite{yamamoto2014stereotactic} and several ongoing randomized trials explore SRS vs WBRT for up to 20 BMs.\cite{suh2020current}

SRS, as opposed to WBRT, delivers a highly conformal dose to the targets in a single fraction (or a few fractions in case of fractionated radiosurgery). This requires high precision in both target localization and treatment delivery.
The concept was first described by Leksell in the 1950s,\cite{leksell1951stereotaxic} leading up to the creation of the Leksell Gamma Knife (Elekta, Sweden), in which a large number of \ce{^{60}Co} sources irradiate a small target volume with high precision thanks to a metal frame attached to the patient's skull.\cite{ICRU_91}
Less invasive fixation devices have later been developed and are now used to deliver SRS treatments also with conventional C-arm linear accelerators (linacs).\cite{gill1991relocatable, ICRU_91} 
One benefit of such linacs is that they are widely available compared to dedicated systems such as Gamma Knife or CyberKnife (Accuray, USA),\cite{fiagbedzi2023radiotherapy} which is important given the growing population of patients that can benefit from these treatments.

Linac-based SRS for multiple BMs has traditionally used a separate isocenter per lesion, leading to prohibitively long treatment times for more than a few metastases.\cite{clark2010feasibility, suh2020current}
To mitigate this, single-isocenter techniques have been developed where multiple targets are treated simultaneously,\cite{vergalasova2019multi} employing either dynamic conformal arc therapy (DCAT),\cite{huang2014radiosurgery, gevaert2016evaluation} where the multi-leaf collimator (MLC) is conformed to the geometrical projection of the targets onto a plane orthogonal to the beam direction called beam's eye view (BEV), or volumetric modulated arc therapy (VMAT),\cite{kang2010, clark2010feasibility, ohira2018hyperarc} where the plan generation algorithm is allowed to position the MLC leaves more freely relative to the BEV projections of the targets.

Simultaneous treatment of several metastases decreases the delivery time,
which can improve availability and cost-effectiveness of the treatment and make it more tolerable for the patient.
However, plan quality may be negatively affected compared to the traditional non-simultaneous treatment.\cite{wu2016optimization}
When multiple metastases are treated simultaneously, the MLC must often be opened in a way that exposes healthy tissues between the metastases. 
This issue is called the island blocking problem.\cite{kang2010}
For a given set of metastases, the occurrence of island blocking can be minimized by collimator angle optimization.\citep{kang2010, wu2016optimization, macdonald2018dynamic, ohira2018hyperarc, battinelli2021collimator} 
However, when there are four or more metastases, even the optimal collimator angle will often lead to an
undesirable amount of exposed normal tissue between targets.

A natural option beyond collimator angle optimization is arranging the targets into two or more subsets which are then treated by separate arc beams
with individually selected collimator angles,
see Figure~\ref{fig:partitioning_example_BEV}.  
This can be understood as a middle-ground between the two extremes: treating one target at a time or treating all targets simultaneously.

Previous approaches for grouping of targets have been suggested in the literature and are used in clinical practice.
For example, 
Kang et al.~\cite{kang2010} partitioned the targets into two groups by exhaustively searching through all such partitions, ranking them with a heuristic objective function,
Chang et al.~\cite{chang2018restricted} grouped the targets into multiple clusters, with limits on the number of targets and the inter-target distance within each cluster, 
and
Palmiero et al.~\cite{palmiero2021management} split the targets into two groups based on location (left vs right hemisphere).
Moreover, a Brainlab Elements (Brainlab, Germany) algorithm optionally divides targets into two groups per couch angle. \cite{gevaert2016evaluation} 
However, all of these methods have potential drawbacks.
They either group the targets into at most two groups, which can limit the usefulness of the methods for patients with many metastases, or use a heuristic objective function, which can lead to suboptimal solutions that fail to minimize island blocking.

In this study, we propose a more general solution to the problem of target partitioning.
The proposed algorithm divides the targets into any number of groups and the groups are determined in a way that minimizes island blocking (or some other cost function) while also respecting user-defined limits on the number of arc beams.
The number of feasible solutions increases rapidly with the number of targets.
However, using a recursive approach, we solve the problem to optimality in reasonable time for up to 20 targets.

% [FIGURE 1]
\begin{figure}[ht]
\begin{center}
\includegraphics[width=\linewidth]{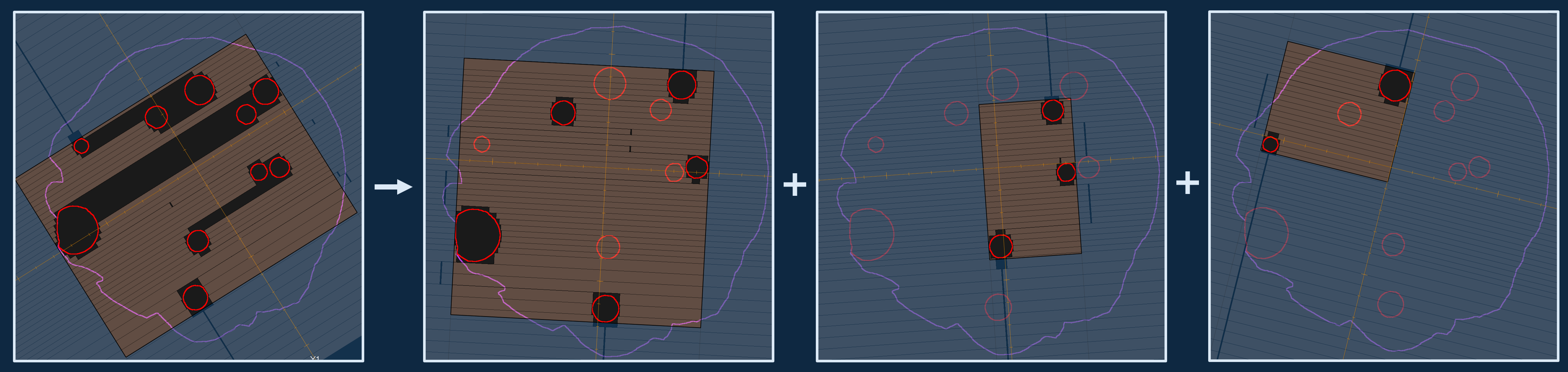}
\captionv{16}{}
{
    \label{fig:partitioning_example_BEV}
    Illustration of the target partitioning approach.
    The island blocking problem occurs when targets line up in the direction of leaf travel, leading to exposure of normal tissue between the targets, as shown in the leftmost aperture. By partitioning the set of targets into subsets, we may avoid the island blocking problem completely, as in the three apertures to the right.
}
\end{center}
\end{figure}

\section{Methods}
In this section, we first devise an algorithm for partitioning a number of metastases into subsets to be treated by separate arcs in a way that minimizes the amount of island blocking while keeping the number of arcs within given limits. 
We then specify experiments for evaluating the proposed algorithm with respect to quality measures such as conformity index, gradient index, and efficiency index, and compare it to simpler alternatives.

\subsection{Proposed algorithm for target partitioning}
\label{sec:proposed_algorithm}
A partition of a set is a division of its elements into non-empty and non-overlapping subsets. 
In this work, the set of interest consists of metastases to treat along some given arc trajectory, and after partitioning, we treat each subset with separate back-and-forth arc passes.

We first look for the optimal partition into exactly $n$ subsets. Later, we generalize the method to also find the optimal $n$.
Let $S$ be a set of metastases to be treated along arc trajectory $a$ and let $\setofpartitions{S}{n}$ be the set of all partitions of $S$ containing exactly $n$ subsets.
Then, our optimization problem amounts to finding the best possible partition in $\setofpartitions{S}{n}$.
We define the cost of a partition $P$ as the sum of the costs of its subsets. With this definition, the optimal cost can be expressed as
\begin{equation}
    \label{eq:optimal_cost}
    f_a(S, n) = \min_{P\in \setofpartitions{S}{n}} \sum_{s \in P} c_a(s)
\end{equation}
where $c_a(s)$ is the cost of a treating subset $s\subseteq S$ along arc trajectory $a$. The function $c_a$ can be selected to penalize various aspects of the subset choice, such as the island blocking area, area of passive leaf gaps, subset size, within-group target distance, and so on. In the present paper, we select $c_a$ to measure the amount of island blocking given an optimal (integer multiple of $1\unit{\degree}$) choice of a static collimator angle. 

We define the amount of island blocking as the BEV area that is exposed when treating all targets simultaneously but not exposed when treating any of the targets individually, e.g., the area between the targets in the leftmost aperture in Figure~\ref{fig:partitioning_example_BEV}. 
For a given collimator angle and subset of targets, we compute this quantity by conforming the MLC to the targets and summing the island blocking area of each individual leaf pair over all control points and leaf pairs. The partitioning algorithm does not make any assumptions regarding the type of MLC used, nor regarding the shapes of the targets. However, with our choice of cost function, both these aspects naturally affect the subset costs and thus the optimal partition.

Evaluating the right-hand side of \eqref{eq:optimal_cost} is equivalent to finding an optimal partition, i.e., the $P$ that minimizes the sum.
This is generally computationally hard since the number of subsets and partitions of $S$ grows quickly with the number of elements.
For instance, there are \num{34105} ways to partition 10 elements into exactly four subsets, while the corresponding number for 20 elements is \num{45232115901}.
However, exact algorithms that avoid complete enumeration of all partitions have been studied before.
We choose to extend a dynamic programming approach\cite{yun1986dynamic, michalak2016hybrid} with a constraint on the number of subsets in the optimal partition, i.e., on the number of arc passes used to treat all metastases in $S$. This allows us to reformulate \eqref{eq:optimal_cost} as a recursion for the non-trivial case $n \geq 2$:
\begin{equation}
    % Modified version of therorem~4 in \cite{michalak2016hybrid} (credited to \cite{yun1986dynamic}) with an additional constraint on the number of subsets in the partition.
    \label{eq:recursion}
    \begin{aligned}
    f_a(S, n) &=
    \begin{cases}
    \displaystyle \min_{s_1, s_2} \quad & c_a(s_1) + f_a(s_2, n-1)\\
    \textrm{s.t.}
    \quad & \{s_1, s_2\} \in \setofpartitions{S}{2} \\  
    \quad &  \abs{s_2} \geq n - 1 \\
    \end{cases}
    \end{aligned}
\end{equation}
where $\abs{\cdot}$ denotes the size of a set. The base case $n = 1$ is trivial since $\setofpartitions{S}{1}$ contains a single partition containing a single element (namely $S$), so $f_a(S, 1) = c_a(S)$.

With formulation~\eqref{eq:recursion}, evaluating $f_a(S, n)$ involves evaluating $f_a(S', n')$ for all valid combinations of $S'\subseteq S$ and $n' \leq n$. 
This convenient fact is exploited by our generalized algorithm, which finds the optimal partitions for multiple arc trajectories simultaneously while respecting constraints on the number of arc passes---both per arc trajectory and in total.
Please refer to Appendix~\ref{appendix:multiple_trajectories} for details.

\subsection{Simulation experiments}
The proposed multi-target partitioning algorithm was evaluated on a number of simulated cases based on anonymized computed tomography (CT) images from a single patient.
In a prestudy, we first investigated how many arcs (arc passes) were needed to eliminate the island blocking problem in cases with up to 20 metastases.
We then incorporated the proposed algorithm into an automatic VMAT planning workflow to evaluate how the number of arcs influence plan quality.
In this main part of the study, we focused on cases with exactly 10 metastases and benchmarked the proposed algorithm against four baseline strategies.
The simulation experiments were run in a research version of RayStation 2023B (RaySearch Laboratories, Sweden).

\subsubsection{Simulated metastases}
For the spatial distribution of brain metastases, we used a simplistic model with randomized spherical targets.
For each case, the gross tumor volume (GTV) radii were sampled independently from a truncated log-normal distribution. 
This choice was inspired by the observation that the size distribution of brain metastases is generally non-normal.\cite{buller2020brain}
We used parameters $\mu = \ln{0.5}$~\unit{\centi\meter} and $\sigma = 0.5$~\unit{\centi\meter} for the underlying normal distribution and truncated to the interval $\mathopen[ 0.1, 1.4 \mathclose]$~\unit{\centi\meter} by rejection sampling.
Given the sampled GTV radii, we then sampled the corresponding sphere center coordinates iteratively, rejecting any choice leading to overlapping planning target volumes (PTVs).
The PTV contours were defined by adding a 1~\unit{\milli\meter} GTV-PTV margin and intersecting the resulting spheres with the brain region of interest (ROI).

A full list of PTV radii, volumes, and coordinates relative the isocenter for the $20\times10$ simulated metastases used in the main study is provided as supplementary material.
In summary, PTV volumes ranged from $0.05~\unit{\centi\meter\cubed}$ to $11.5~\unit{\centi\meter\cubed}$ with an average of $1.5~\unit{\centi\meter\cubed}$, and the distance to isocenter ranged from $0.8~\unit{\centi\meter}$ to $8.9~\unit{\centi\meter}$ with an average of $4.7~\unit{\centi\meter}$.

Some guidelines for SRS treatments of brain metastases suggest lower prescription doses and/or increased fractionation for larger metastases.\cite{gondi2022ASTROguidelines}
However, to simplify plan evaluation in the simulation study, we assumed a prescription of 21~\unit{\gray} near-minimum dose in one fraction to all metastases, regardless of size. 

\subsubsection{Beam setup and arc trajectories}
\label{sec:arc_trajectories}
Treatment plans were generated for a 6~\unit{\mega\volt} flattening filter free (FFF) beam on a Varian TrueBeam STx LINAC equipped with a HD120 MLC (Varian Medical Systems, USA), for which the 32 most central MLC leaf pairs appear 2.5~\unit{\milli\meter} wide projected onto the isocenter plane. The maximum dose rate was set to 1400~\unit{MU\per\minute}.

For each plan, we defined a single isocenter in the unweighted center of the individual PTV's centers of mass and created three arcs above the head at couch angles \ang{90}, \ang{30}, and \ang{330}
(IEC~61217). Each arc had a gantry angle span of \ang{176}, covering the ranges [\ang{182}, \ang{358}] or [\ang{2}, \ang{178}] with a \ang{2} control point spacing.
Note that the partitioning algorithm can be used in combination with any number of arc trajectories.

In RayStation, it is possible to restrict which targets to treat per beam and to specify their associated margins. This effectively defines bounds for the MLC leaf positions relative to the BEV projections of the targets. 
In our template beam setup, all PTVs were initially specified to be treated along all three arc trajectories with uniform 0~\unit{\milli\meter} margins.
As discussed in Section~\ref{sec:vmat_optimization_strategy}, the margins were increased to 1~\unit{\milli\meter} later in the  process.

\subsubsection{Prestudy on the number of arcs required to eliminate island blocking}
As a prestudy, we simulated $10\times 20$ patient cases with $1$--$20$ metastases and applied the arc trajectory setup described in section~\ref{sec:arc_trajectories}. 
For each case, we ran the proposed partitioning algorithm without limits on the number of arcs and recorded the minimum total number of arcs needed to eliminate the island blocking problem. 
We considered the island blocking problem eliminated when, for each arc, the island blocking area summed to less than $0.5~\unit{\centi\meter\squared}$ over all control points, given an optimal (integer multiple of $1\unit{\degree}$) static collimator angle.

\subsubsection{Target partitioning and baseline algorithms}
The proposed algorithm was run 13 times with different settings for each of the 20 simulated patient cases.
Each run used a different equality constraint on the total number of arcs, $N$.
Based on the prestudy, $N\approx 10$ should often suffice to eliminate island blocking and thus saturate our objective function \eqref{eq:optimal_cost}.
We decided to run the simulation experiments for $N$ in the range 3--14, but we also included $N=30$ to cover the extreme case of treating one target at a time.
After running the algorithm and updating the beam setup based on the result, we optimized all collimator angles with the standard DCAT collimator angle optimization in RayStation,\cite{battinelli2021collimator} which finds the collimator angle with the minimum exposed non-target area summed over the control points.

For comparison, we evaluated three additional target partitioning strategies for which the target subsets were determined by either clustering, anticlustering, or random sampling. For these baseline strategies, the number of arcs were always distributed as evenly as possible across the couch angles, e.g., as $2+2+3$ when $N = 7$. 
The clustering strategy used an in-house implementation of $k$-means clustering to group nearby targets into subsets. 
The anticlustering strategy instead created as disperse subsets as possible, using version 0.6.4 of the R-package anticlust\cite{papenberg2021anticlustering} and a distance matrix based on distances between the PTV surfaces.
The random sampling strategy used NumPy 1.24.3 to randomly permute the targets before splitting them into evenly sized subsets.
Similarly as for the proposed algorithm, the standard collimator angle optimization in RayStation was applied after partitioning for all three baseline partitioning strategies.

For completeness, we also included a baseline strategy that created duplicate beams without dividing the targets into subsets, allowing all arcs to treat all targets.
Again, the number of arcs were distributed as evenly as possible across the three couch angles. Standard collimator angle optimization would have given identical collimator angles for the duplicated arcs. To avoid this, we used an ad hoc sequence of collimator angles defined as
\begin{equation}
    \label{eq:duplicates_angles}
    \left( \frac{ (2 k - 1)}{\floor{(n - 1)/2} + 1} \times \ang{45} \right)_{k = 1}^{n}
\end{equation}
for a couch angle with $n$ duplicated arcs, e.g., $(\ang{22.5}, \ang{67.5}, \ang{112.5})$ for $n=3$. 
% In one single case, we replaced $\ang{45}$ with $\ang{-45}$ to circumvent an unexpected premature termination of the optimization.

\subsubsection{Dose-based optimization}
\label{sec:vmat_optimization_strategy}
The choice of optimization objectives has a large impact on the optimized plan. For example, limiting the max dose within the target negatively affects the dose gradient, \cite{morrison2016single, vergalasova2019multi} but institution-dependent plan criteria might still limit the max dose to some percentage of the prescription.\cite{ballangrud2018institutional}
In this study, we used the objectives listed in Table~\ref{tab:objective_functions}, including a max dose objective at 140\% of the prescription dose.

% [TABLE 1, tab:objective_functions]
\begin{table}[ht]
\begin{center}
\captionv{16}{}
{RayStation dose-based objective functions used in the simulation experiments. Numbers in parenthesis show adjustments for the clinical case study.}
\label{tab:objective_functions}
\vspace*{2ex}
\footnotesize
\begin{tabular}{lll}
\toprule
ROI & Objective & Weight \\
\midrule
$\text{PTV}_i$ & Min DVH 100\% (101\%) of prescription to 98\% (99\%) volume & $1000^\dag$ \\
$\text{PTV}_i$ & Min dose 100\% of prescription & 100 \\
$\text{PTV}_i$ & Max dose 140\% (130\%) of prescription & 100 \\
$\text{PTV}_i$ & Max DVH 130\% (120\%) of prescription to 5\% volume & 5 \\
\multirow[t]{2}{*}{$\text{PTV}_i$ 0--10~\unit{\milli\meter} ring} & Dose fall-off [H]100\% of prescription [L]50\% of prescription, & 10 \\
& Low dose distance 0.30~\unit{\centi\meter} &  \\
\multirow[t]{2}{*}{External} & Dose fall-off [H]100\% of prescription [L]25\% of prescription, & 10 \\
& Low dose distance 1.00~\unit{\centi\meter} & \\
$\text{OAR}_j$ & Max dose 80\% of acceptance level & 0 (10) \\
\bottomrule
\end{tabular}
\begin{tabular}{c}
    $^\dag$Also included as a constraint in the second optimization phase.\\
\end{tabular}
\end{center}
\end{table}

A scripted VMAT optimization procedure was run for all competing algorithms.
As a first step, all BEV treat margins were increased to 1~\unit{\milli\meter} since an initial experiment had shown that 0~\unit{\milli\meter} margins were too small for the smallest targets.
In the first optimization phase, $2\times40$ iterations were run on a uniform 1.5~mm dose grid, including 7 iterations before segment conversion.
In the second optimization phase, the dose grid resolution was decreased to 1.0~mm and a number of constraints were added to the problem (see Table~\ref{tab:objective_functions}) before running another 40 iterations.
In this phase, aperture shapes were kept fixed and only segment MUs were optimized.
This ensured agreement between the optimization dose and final dose, and, in combination with the constraints, proper fulfillment of the prescriptions. 
Final dose was computed with the Collapsed Cone algorithm.

% \clearpage
\subsubsection{Plan evaluation metrics}
\label{sec:plan_evaluation}
The optimized plans were evaluated with respect to several metrics chosen based on ICRU Report~91.\cite{ICRU_91}
We computed the near-minimum ($\text{D}_{98\%}$), median ($\text{D}_{50\%}$), and near-maximum ($\text{D}_{2\%}$) dose to each individual PTV.
Furthermore, we used the Paddick conformity index (PCI) to evaluate conformity and the gradient index (GI) to evaluate dose fall-off.

PCI is defined as:
\begin{equation}
    \text{PCI} = 
    \frac{\text{TV}^2_{\text{PIV}}}{\text{TV} \times \text{PIV}},
    \label{eq:paddick_conformity_index}
\end{equation}
where $\text{TV}$ is the target volume, $\text{PIV}$ the prescription isodose volume, and $\text{TV}_{\text{PIV}}$ the target volume within the prescription isodose.\cite{paddick2000conformityIndex} 

GI is defined as:
\begin{equation}
    \text{GI} = 
    \frac{\text{PIV}_{\text{half}}}{\text{PIV}}
    \label{eq:gradient_index},
\end{equation}
where $\text{PIV}_{\text{half}}$ is the volume receiving at least half the prescription dose.\cite{paddick2006gradientIndex}

Sometimes in multi-target cases, PCI and GI are reported per target and not per plan.
However, it is not clear how to compute these indices when the relevant isodose surfaces do not form isolated islands around each target. This is especially problematic for GI since $\text{PIV}_{\text{half}}$ is a larger volume than $\text{PIV}$.\cite{ICRU_91}
Several strategies have been proposed for attributing only part of the isodose volumes to each target in such cases. \cite{cui2020retrospective, desai2022estimate}
In this work, we report PCI per individual PTV, only considering the PIV within a 5~\unit{\milli\meter} expansion around each target, after first verifying that the PIV never extended more than 5~\unit{\milli\meter} away from the PTV union in any of the analyzed plans. 
For GI, we report one number per plan, computed for the PTV union.
This approach was straightforward since all targets shared the same prescription dose level.

As complementary metrics, we included the volume of the brain, excluding PTVs, receiving at least 12~\unit{\gray} ($\vtwelve$), and the global efficiency index (G$\eta$), defined as the ratio between the integral dose within the target volume and the integral dose within the 12~\unit{\gray} isodose volume.\cite{dimitriadis2018novel}
Finally, we also included the total number of monitor units (MUs) and the total beam-on time as estimated by RayStation.

\subsubsection{Statistical analysis}
\label{sec:method_statistics}
Two-tailed Wilcoxon signed-rank tests with paired data were used to compare the proposed algorithm against the baselines.
Separate tests were run using SciPy 1.10.1 for each combination of algorithm, metric, and total number of arcs.
To mitigate the multiple comparisons problem, tests differing only in the total number of arcs were grouped into families for which the Holm-Bonferroni method\cite{holm1979simple} was applied. This procedure keeps the family-wise error rate (probability of at least one false positive) below $\alpha =0.05$ by using corrected $p$-values in the range [$\alpha/m$, $\alpha$], where $m$ is the number of tests in the family.
However, we are still left with 40 families of tests, meaning that at least some false positives in some families must be expected.
To complement each significance test, we generated a box plot over the pairwise differences to illustrate the sign, magnitude and variability of the effect.

\subsection{Clinical cases}
To further verify the proposed algorithm, we performed a retrospective planning study on six anonymized clinical cases (one re-treatment) for five patients treated at the Iridium Netwerk.
Due to the de-identified nature of the data, the need for ethics review was waived for this retrospective study.
The six cases had 5, 6, 6, 6, 9 and 10 metastases respectively.
In summary, PTV volumes ranged from $0.2~\unit{\centi\meter\cubed}$ to $12.9~\unit{\centi\meter\cubed}$ with an average of $1.6~\unit{\centi\meter\cubed}$, and the distance to isocenter ranged from $1.4~\unit{\centi\meter}$ to $7.4~\unit{\centi\meter}$ with an average of $4.5~\unit{\centi\meter}$.
A full list is provided in the supplementary material.

The prescription dose level and fractionation varied between 
from case to case, but not between targets within each case. 
Prescriptions included:
1x18~\unit{\gray},
1x21~\unit{\gray},
3x8~\unit{\gray},
3x9~\unit{\gray},
and
5x6~\unit{\gray} to 99\unit{\percent} volume of each PTV.
Further clinical goals were to keep the PTV \doseatccvolume{0.1} below 140\unit{\percent} of the prescription dose level, and for organs at risk (OARs) to keep the \doseatccvolume{0.1} below acceptance levels that varied with the number of fractions as follows:
1.5~\unit{\gray}/1~fx,
2.1~\unit{\gray}/3~fx,
or
5~\unit{\gray}/5~fx 
for the lenses;
8~\unit{\gray}/1~fx,
15~\unit{\gray}/3~fx,
or
22.5~\unit{\gray}/5~fx 
for the optic chiasm, optic nerves, and eyes; 
and
10~\unit{\gray}/1~fx,
18~\unit{\gray}/3~fx,
or
23~\unit{\gray}/5~fx 
for the brainstem.

Analysis of the clinical plans revealed that 5--8 arcs per case were used, each arc treating 2--5 targets (average 3.4) over a gantry angle span of 130\unit{\degree}--178\unit{\degree} or 348\unit{\degree} (average 178\unit{\degree}). The arcs in each plan were configured to treat each target from 2--4 couch angles (average 2.9) over a total gantry angle span of 346\unit{\degree}--864\unit{\degree} (average 527\unit{\degree}).
The couch angles, gantry angle spans, targets to treat per arc, and collimator angles had been selected manually with the aim to minimize island blocking while also considering OARs and the fact that each PTV must be treated from sufficiently many directions. 
This beam configuration process generally requires at least 30 minutes of trial-and-error per case even for an experienced planner. 

We generated two alternative plans for each case: 
one using the beam configuration from the clinical plan and one using a beam configuration generated with the proposed algorithm.
For the latter, we reused the three arc trajectories from the simulation experiments, see Section~\ref{sec:arc_trajectories}, and constrained the number of arcs to a number in 5--8 chosen to approximately match the total gantry angle span of the corresponding clinical plan.
Both new plans were optimized with identical objective functions, adapted from the simulation study as indicated in Table~\ref{tab:objective_functions}, including max dose objectives on the OARs for which max \doseatccvolume{0.1} acceptance levels are listed above.
We used the same VMAT optimization strategy as outlined in Section~\ref{sec:vmat_optimization_strategy} but with a 1.2~\unit{\milli\meter} dose grid for both optimization and evaluation to match the clinical plans.

The clinical plans had been generated in a clinical version of RayStation 2024A (RaySearch Laboratories, Sweden), and the reoptimized plans were generated in a research version of RayStation 2024B (RaySearch Laboratories, Sweden).
For evaluation, all doses were computed on a 1.2~\unit{\milli\meter} dose grid with the Collapsed Cone algorithm in RayStation 2024B.

% \clearpage
\section{Results}
\subsection{Prestudy}
With a target partitioning approach, the number of arcs needed to fully eliminate island blocking naturally increases with the number of targets. 
The purpose of the prestudy was to investigate this correlation given our beam setup
and method for simulating metastases.

As seen in Figure~\ref{fig:perfect_partitions}, the required number of arcs increased with the number of targets in a slightly sublinear way.
For example, most cases with 5 targets required a total of 6 arcs. 
By design, each target was treated by exactly 3 arcs, and thus each arc treated $(3 \times 5) / 6 = 2.5~\unit{targets}$ on average in these cases.
Similarly, each arc treated about 3 targets on average in cases with 10 targets, and about 4 targets on average in cases with 20 targets.

% [FIGURE] figure_2.pdf
\begin{figure}[ht]
\begin{center}
\includegraphics[width=0.95\linewidth]{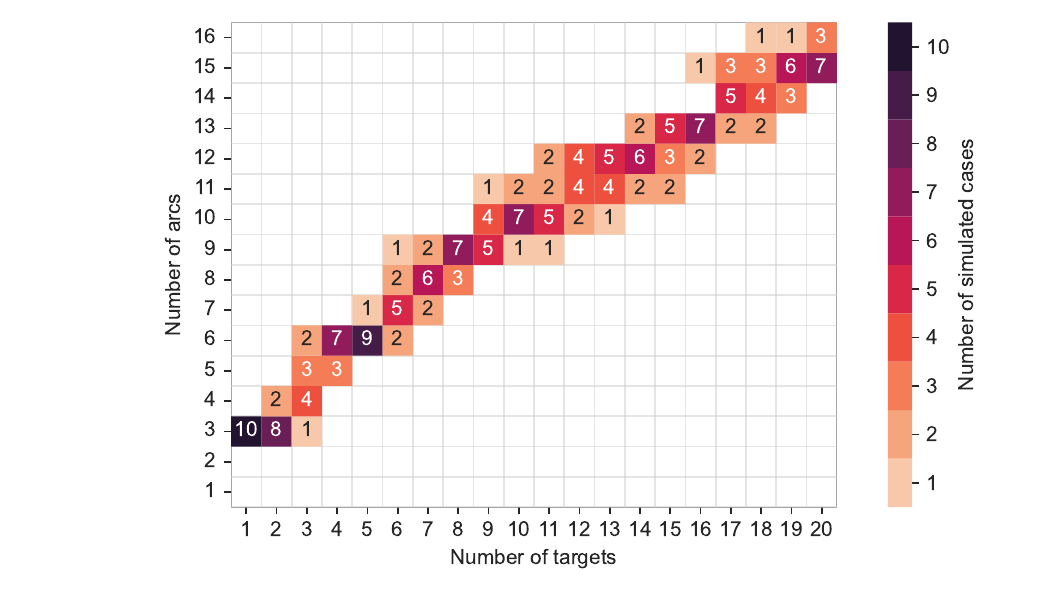}
\captionv{16}{}
{
Summary of the 200 prestudy experiments. The values on the y-axis represent the minimum number of arcs (summed over the three couch angles) required to fully eliminate island blocking using a target partitioning approach.
\label{fig:perfect_partitions} 
}
\end{center}
\end{figure}

The algorithm runtime increased with the number of targets, staying below 5~\unit{\minute} per case for up to 15~targets, then increasing more rapidly to 10~\unit{\minute} for 17~targets and 84~\unit{\minute} for 20~targets.

% \clearpage
\subsection{Simulation experiments}
Table~\ref{tab:proposed_mean} shows averaged evaluation metrics for the proposed target partitioning approach. As expected, the target partitioning cost \eqref{eq:optimal_cost} decreased with increasing number of arcs. The average PTV dose-volume histogram (DVH) metrics ($\text{D}_{98\%}$, $\text{D}_{50\%}$, $\text{D}_{2\%}$) and conformity index (PCI) did not vary much, but the dose gradient clearly improved with increasing number of arcs, as indicated by
the decrease in GI and $\vtwelve$ as well as the increase in G$\eta$.
These improvements also correlated with decreasing objective values. 
On the downside, the MUs and beam-on time increased steadily with the number of arcs.

% [TABLE] tab:proposed_mean
\begin{table}[ht]
\begin{center}
\captionv{16}{}
{
Overview of simulation experiment results for the proposed algorithm.
P.~obj.~is the target partitioning cost function \eqref{eq:optimal_cost} summed over the three arc trajectories.
D.~obj.~is the total dose-based objective value relating to the objective functions listed in Table~\ref{tab:objective_functions}. 
$\text{D}_{98\%}$, $\text{D}_{50\%}$, $\text{D}_{2\%}$, and PCI are averaged over the $20\times10$ simulated PTVs, whereas the other metrics are averaged over the 20 plans.
$\vtwelve$ is the volume of the brain, excluding PTVs, receiving at least $12~\unit{\gray}$. Time refers to the beam-on time as estimated by RayStation.
}
\label{tab:proposed_mean}
\vspace*{2ex}
\sisetup{
    mode=text,
    group-digits=integer, % https://tex.stackexchange.com/a/561368
    group-minimum-digits=5,
    tight-spacing=true,
    round-mode = places,
}
\footnotesize
\begin{tabular}
{
S[table-format=2.0, round-precision=0] % Arcs
|
S[table-format=4.2, round-precision=2] % P.~obj.
S[table-format=1.3, round-precision=3] % D.~obj.
S[table-format=2.1, round-precision=1] % $D_{98\%}$
S[table-format=2.1, round-precision=1] % $D_{50\%}$
S[table-format=2.1, round-precision=1] % $D_{2\%}$
S[table-format=1.3, round-precision=3] % PCI
S[table-format=1.2, round-precision=2] % GI
S[table-format=1.3, round-precision=3] % G$\eta$
S[table-format=2.1, round-precision=1] % $V_{12~\unit{\gray}}$
S[table-format=5.0, round-precision=0] % MU
S[table-format=2.2, round-precision=2] % Time
}
\toprule
{Arcs} & {P.~obj.} & {D.~obj.} & {$\text{D}_{98\%}$} & {$\text{D}_{50\%}$} & {$\text{D}_{2\%}$} & {PCI} & {GI} & {G$\eta$} & {$\vtwelve$} & {MU} & {Time} \\
\midrule
3 & 2251.848057 & 0.479547 & 21.002775 & 24.396566 & 27.475412 & 0.793307 & 5.850679 & 0.322274 & 48.568561 & 6801.200783 & 4.935114 \\
4 & 1462.027142 & 0.233475 & 21.029890 & 24.747050 & 27.778406 & 0.804667 & 4.411534 & 0.386895 & 36.056555 & 8565.054100 & 6.286473 \\
5 & 766.202759 & 0.134146 & 21.056098 & 24.960041 & 27.866611 & 0.808544 & 3.773044 & 0.425625 & 30.191887 & 9436.711918 & 7.013618 \\
6 & 160.336725 & 0.100191 & 21.084817 & 24.978069 & 27.755451 & 0.809828 & 3.516283 & 0.441463 & 27.952402 & 9846.139349 & 7.401002 \\
7 & 90.051384 & 0.081415 & 21.088957 & 25.051437 & 27.817857 & 0.810664 & 3.406650 & 0.450073 & 26.894159 & 11612.671203 & 8.676602 \\
8 & 33.669940 & 0.073511 & 21.101506 & 25.061617 & 27.807400 & 0.810807 & 3.350159 & 0.454715 & 26.295845 & 13023.046951 & 9.684422 \\
9 & 1.862410 & 0.066921 & 21.109202 & 25.102837 & 27.816386 & 0.810399 & 3.287688 & 0.459114 & 25.786958 & 14358.970470 & 10.647903 \\
10 & 0.183214 & 0.067423 & 21.116174 & 25.111297 & 27.830543 & 0.809150 & 3.294670 & 0.458742 & 25.843501 & 16001.829331 & 11.844658 \\
11 & 0.005056 & 0.063933 & 21.122953 & 25.131256 & 27.827451 & 0.810408 & 3.265507 & 0.461341 & 25.586464 & 17397.433127 & 12.928882 \\
12 & 0.000000 & 0.063311 & 21.119004 & 25.118154 & 27.807533 & 0.810772 & 3.258448 & 0.461470 & 25.547610 & 19123.249053 & 14.198505 \\
13 & 0.000000 & 0.061977 & 21.123057 & 25.124677 & 27.779541 & 0.810490 & 3.249102 & 0.462668 & 25.381914 & 20658.153230 & 15.329342 \\
14 & 0.000000 & 0.060291 & 21.139458 & 25.140728 & 27.779467 & 0.809760 & 3.225293 & 0.463871 & 25.263751 & 22340.945852 & 16.574025 \\
30 & 0.000000 & 0.044803 & 21.193074 & 25.187327 & 27.594826 & 0.810578 & 3.032373 & 0.478916 & 23.694971 & 47265.006718 & 35.073329 \\
\bottomrule
{-} & {\unit{\centi\meter\squared}} & {-} & {\unit{\gray}} & {\unit{\gray}} & {\unit{\gray}} & {-} & {-} & {-} & {\unit{\centi\meter\cubed}} & {-} & {\unit{\minute}}
\end{tabular}
\end{center}
\end{table}

The dose-based objective values are plotted in Figure~\ref{fig:simulation_box_plots}a.
Note that for 3 and 30 arcs, the four target partitioning approaches are equivalent and correspond to treating all targets simultaneously (3 arcs) or all targets individually (30 arcs).
This explains the identical results. For 3 arcs, the duplicates strategy is also equivalent except for the collimator angle, which was set to a constant $45\unit{\degree}$ following~\eqref{eq:duplicates_angles} rather than optimized. Evidently, optimizing the collimator angle was beneficial.

% [FIGURE] figure_3.pdf
\begin{figure}[ht]
\begin{center}
\includegraphics[width=\linewidth]{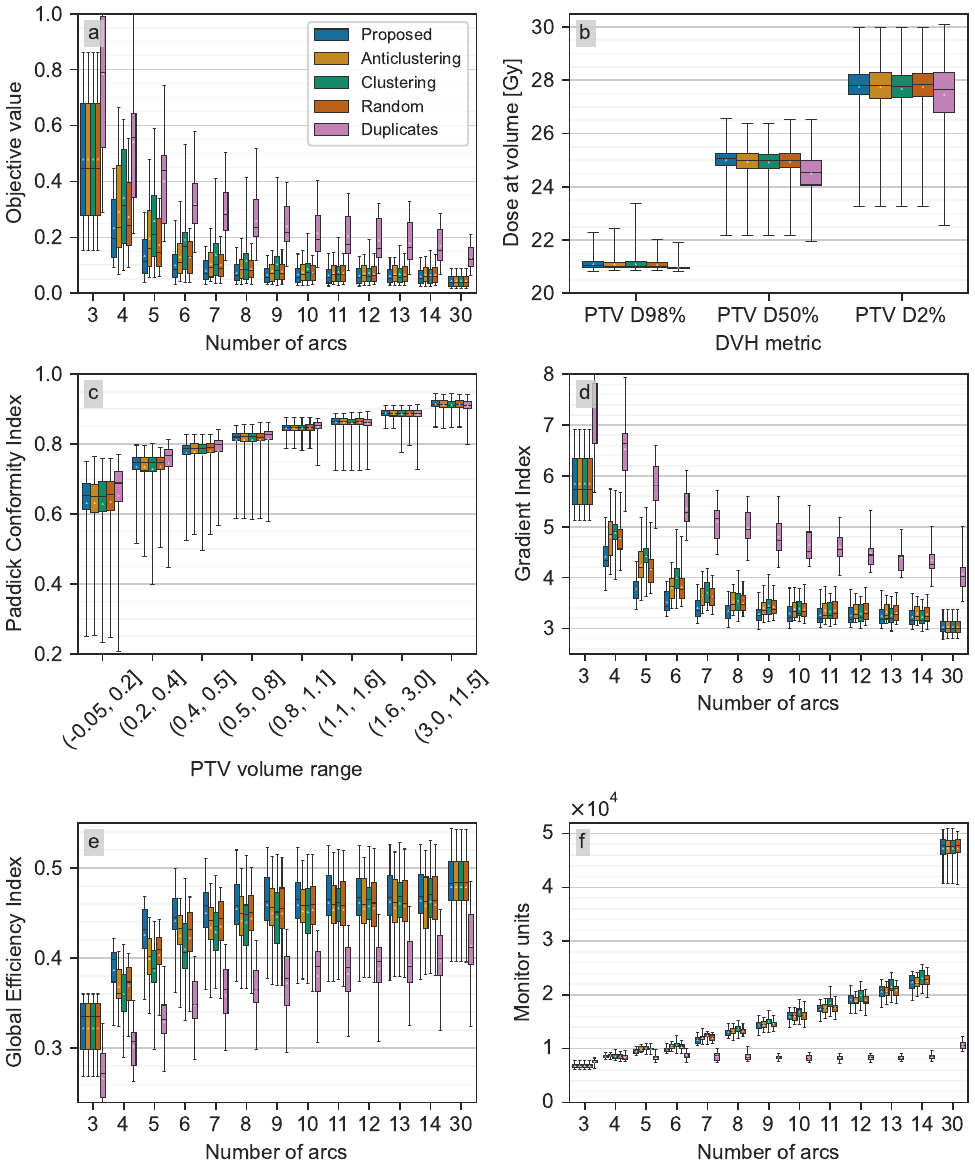}
\vspace*{-5ex}
\captionv{16}{}
{
Plan quality comparison between the proposed algorithm and the baselines.
White triangles indicate mean values.
Note that objective value refers to the dose-based objectives used in the VMAT optimization, and that PCI and the PTV DVH metrics are computed per individual PTV (10 per plan) whereas the other metrics are computed per plan.
\label{fig:simulation_box_plots} 
}
\end{center}
\end{figure}

The box plot in Figure~\ref{fig:simulation_box_plots}a reveals visibly lower (better) objective values for the proposed algorithm compared to all other algorithms, at least for 4--9 arcs, and the statistical analysis (details in the supplementary material) reveals significantly better results also for 10 or more arcs, although with diminishing magnitudes of the median pairwise differences.
The duplicates baseline consistently performed the worst in terms of objective value and did not beat the proposed algorithm in any single pairwise comparison.

As illustrated in 
Figures~\ref{fig:simulation_box_plots}b--c, all plans achieved similar target coverage and conformity.
Some variability is expected given the variability in PTV volumes across patients.
However, this variability is canceled out in the pairwise differences. The median pairwise differences in {$\text{D}_{98\%}$}, {$\text{D}_{50\%}$} and {$\text{D}_{2\%}$} were negligibly small between algorithms, except maybe the median 0.5~\unit{\gray} decrease in $\text{D}_{50\unit{\percent}}$ for the duplicates baseline.
PCI increased slightly with the number of arcs. 
There were some significant PCI differences in favor of the proposed algorithm for plans with few arcs and some in favor of the duplicates baseline for plans with many arcs, but the magnitudes of the median pairwise differences were at most 0.02.

Given the consistency in dose conformity across all plans, the significant differences in objective value are better explained by differences in the dose gradient. 
Figures~\ref{fig:simulation_box_plots}d--e
reveal that both GI and G$\eta$ improved considerably with increasing number of arcs for all algorithms.
The proposed target partitioning algorithm performed the best, in particular for 4--9 arcs, with median pairwise
0.1--0.6 reductions in GI, 
1--3 percentage point increases in G$\eta$, and 
1--6~\unit{\centi\meter\cubed} reductions in $\vtwelve$ 
compared to the other partitioning approaches.
There were also (with some exceptions) significant but small improvements for larger number of arcs.
Compared to the duplicates baseline, the proposed method achieved large and significant improvements, with median pairwise 
1.0--2.2 reductions in GI, 
5--9 percentage point increases in G$\eta$, and
8--15~\unit{\centi\meter\cubed} reductions in $\text{V}_{12~\unit{ \gray}}$.

The trends for GI, G$\eta$, and $\text{V}_{12~\unit{ \gray}}$ are consistent with the trends observed for the objective values in Figure~\ref{fig:simulation_box_plots}a.
Although the improvements continued beyond that point, the box plots for the proposed algorithm show a visible plateau in plan quality around 10 arcs, which corresponds well to the average number of arcs needed to eliminate the island blocking problem for cases with 10 targets, see Figure~\ref{fig:perfect_partitions} and Table~\ref{tab:proposed_mean}.
Figure~\ref{fig:2D_dose_square} visually exemplifies how the dose gradient improved up until said plateau.

% [FIGURE] figure_4.png
\begin{figure}[ht]
\begin{center}
\includegraphics[width=\linewidth]{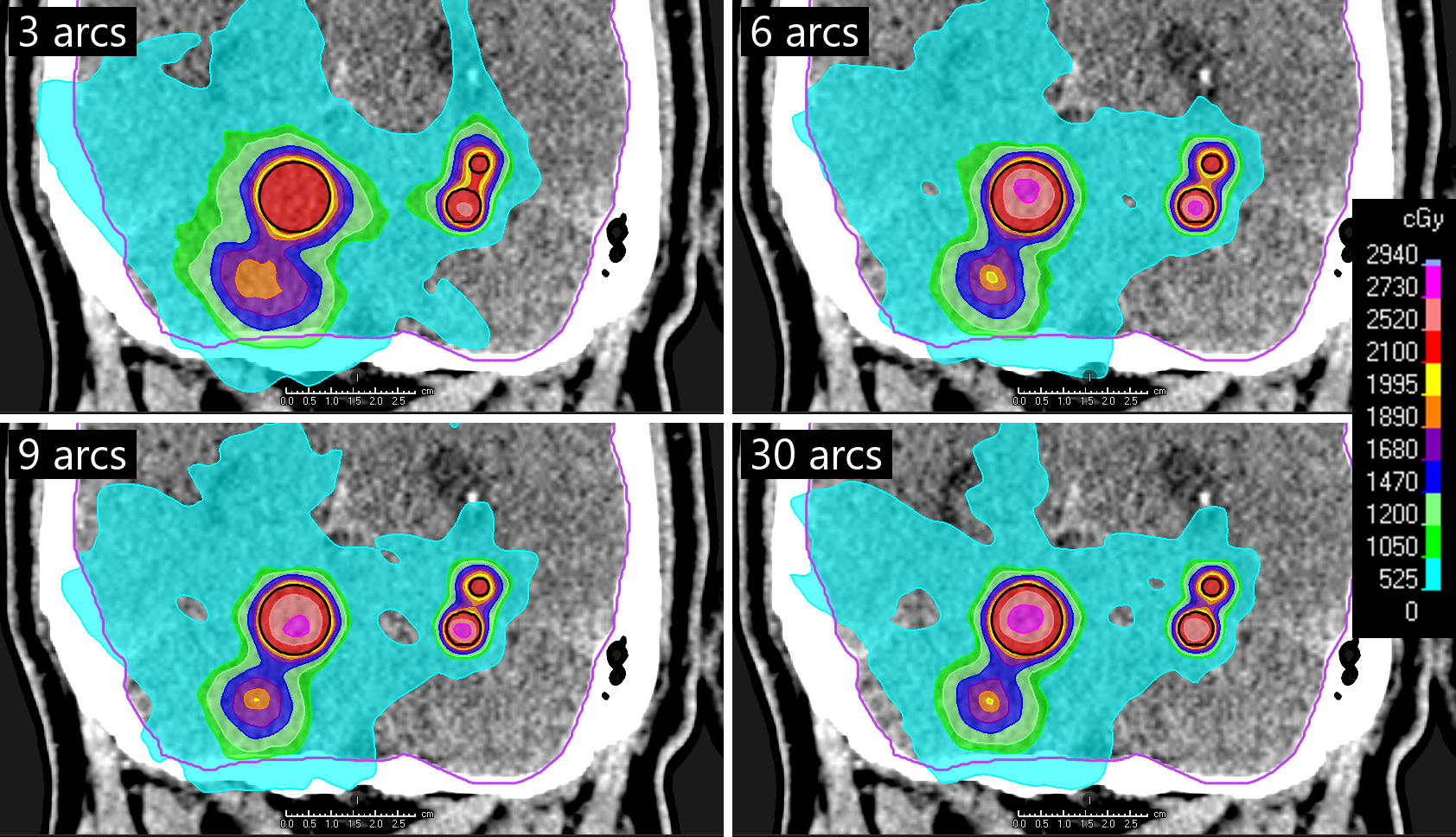}
\captionv{16}{}
{
Delivery time trade-off.
This example illustrates how the dose gradient varied with the number of arcs for the proposed algorithm in one of the simulated cases.
PTV contours are shown in black.
Note how the dose gradient clearly improves between 3, 6, and 9 arcs, whereas the difference between 9 and 30 arcs is small. Since the delivery time is several times longer for 30 arcs, the 9-arc solution is much preferred in clinical practice.
%The dose slices were taken in the coronal direction 1.39~cm posterior of the isocenter for patient~1.
\label{fig:2D_dose_square} 
}
\end{center}
\end{figure}

One major difference between the four target partitioning approaches and the duplicates baseline is illustrated in Figure~\ref{fig:simulation_box_plots}f.
For the partitioning approaches, the MUs increased steadily with the number of arcs, while the MUs stayed relatively constant for the duplicates baseline.
Note that the number of MUs is strongly correlated with the beam-on time and is thus an important contributor to the overall treatment delivery time.

\FloatBarrier
\subsection{Clinical cases}
Figure~\ref{fig:case_study} summarizes the results of the retrospective planning study on six clinical cases.
The number of cases is too low for statistical analysis, but large enough to give a qualitative indication of the performance of the proposed algorithm in realistic cases.

% [FIGURE] figure_5.pdf
\begin{figure}[ht]
\begin{center}
\includegraphics[width=\linewidth]{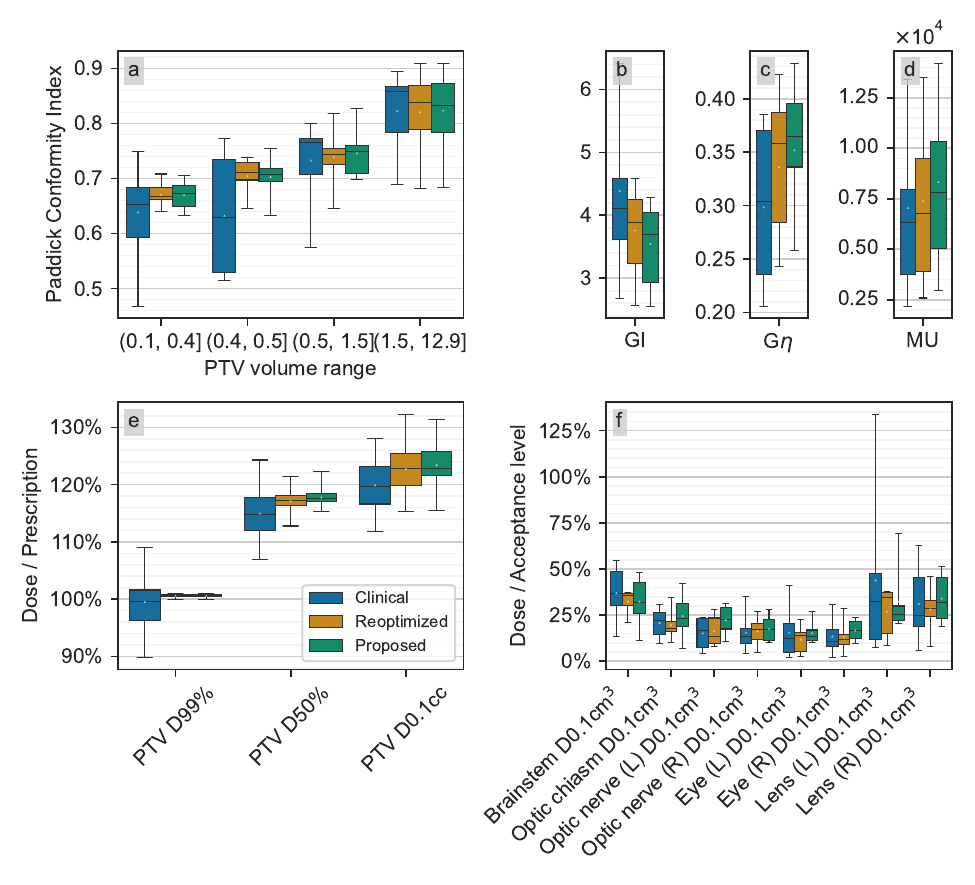}
\captionv{16}{}
{
\label{fig:case_study}
Plan comparison for the six clinical cases included in the retrospective planning study. Since the prescription dose levels and number of fractions varied across cases, the PTV DVH metrics are presented relative the prescription dose level, and the OAR near-max doses (\doseatccvolume{0.1}) are presented relative their fractionation scheme-specific acceptance levels.
}
\end{center}
\end{figure}

The comparison between the clinical plans and the reoptimized plans isolates the differences caused by the updated optimization strategy and dose-based objectives.
For example, the reoptimized plans perfectly fulfilled all PTV~\doseatpercentvolume{99} thanks to the final segment MU optimization with constraints on the prescriptions, as compared to the clinical plans.
Furthermore, the reoptimized plans achieved improved PCI compared to the clinical plans for the smaller PTVs, the GI and G$\eta$ also improved, and because of the explicit inclusion of OAR max dose objectives, no OAR acceptance levels were exceeded.
Overall, the clinical plans and the reoptimized plans were judged to be clinically equivalent.

The comparison between the reoptimized clinical plans and the plans generated with the proposed algorithm isolates differences caused by differing beam configurations.
Even though the proposed algorithm used the same three arc trajectories for all cases and minimum manual intervention (only the choice of total number of arcs per case), the plan quality in terms of PCI, GI, G$\eta$ and clinical goal fulfillment was equivalent or slightly better compared to the reoptimized clinical plans.
The average plan MU increased slightly, but it shall be noted that no explicit MU penalty was included in the optimization.

\section{Discussion}
A sharp dose gradient is of high clinical importance in stereotactic treatments of brain metastases, where for example the volume receiving at least 12~\unit{Gy} has been correlated to the probability of brain radionecrosis.\cite{blonigen2010irradiated, milano2021single}
In our simulated cases with 10 metastases, the dose gradient was much better in terms of average GI (3.0 vs 5.9), G$\eta$ (0.48 vs 0.32), and $\vtwelve$ (24~\unit{\centi\meter\cubed} vs 49~\unit{\centi\meter\cubed}) 
for plans treating one metastasis at a time using 30 arcs, compared to plans treating all metastases simultaneously using 3 arcs.
These improvements required around seven times more MUs.
Note that treating with 30 arcs would in many cases be clinically infeasible as the patient may not tolerate such long treatment times, for example due to back pain or shortness of breath.
However, using our proposed target partitioning algorithm, we achieved almost as good average GI (3.3), G$\eta$ (0.46) and $\vtwelve$ (26~\unit{\centi\meter\cubed}) using only 9 arcs and a little more than twice the MUs of the simultaneous treatment.

All four investigated partitioning approaches resulted in better quality plans than just duplicating the arcs,
but the proposed algorithm, which directly minimizes the amount of island blocking, was the most effective in utilizing a limited budget of arcs, most notably with respect to GI, G$\eta$, and $\vtwelve$.
Combined with the observed plateau in most plan quality metrics coinciding with zero partition cost (Table~\ref{tab:proposed_mean}),
this indicates that the amount of island blocking is indeed a relevant predictor of plan quality for this type of case and optimization strategy.
Such a finding is consistent with previous findings~\cite{kang2010} and the community knowledge.

The relatively poor performance of the baseline approach with duplicate arcs but no subdivision of targets may be surprising.
In fact, with such a setup, the leaf positions are less constrained and could in theory converge towards a partition-like solution if beneficial. 
Such convergence was, however, not observed.
The plausible explanation is that VMAT optimization is a highly non-convex problem that no known method solves to optimality. Gradient-based optimization of aperture shapes will generally converge towards a local optimum dependent on the segment initialization.\cite{unkelbach2015optimization}
One intuitive source of non-convexity is the combination of island blocking and optimization objectives pushing for sharp dose gradients.
In such control points, the affected leaf pairs should ideally shield the normal tissue in between targets by closing over targets on either side. 
Shielding from the left or right might both be acceptable, while a linear interpolation between these solutions is not.

The retrospective planning study on six clinical cases showed that plans generated with the proposed algorithm, even with minimal manual intervention, could achieve similar plan quality as plans using the clinical, hand-crafted beam configurations.
Although the number of cases is too low for significance tests, this indicates that the proposed method generalizes to realistic cases with varying numbers of non-spherical metastases.
The comparison with the clinical plans demonstrated that the objective functions and optimization strategy, not only the beam configuration, are important factors influencing plan quality. 

In this study, we evaluated plan quality based on the physical dose distribution. 
However, previous studies have suggested that prolonged fraction delivery times may reduce tumor control probabilities due to the potential repair of sublethal DNA damage during irradiation. \cite{fowler2004loss, nakano2022radiobiological}
This could warrant a biological perspective on the target partitioning approach, where the dose to each metastasis is primarily delivered in one chunk per couch angle with interruptions both for treating the other metastases and for rotating the couch.
According to Table~\ref{tab:proposed_mean}, average beam-on times remained below 15 minutes for up to 12 arcs. 
Even when accounting for patient setup the time between arcs, the total treatment time is expected to remain in the lower to median range of estimated durations for multi-target brain radiotherapy with systems such as CyberKnife or ZAP-X.\cite{ianiro2023multiple, paddick2023benchmarking, niu2024comparative} 
Nonetheless, future investigations incorporating biologically realistic models that factor in treatment time would provide valuable insights into potential clinical implications.

In our experiments, we have assumed that all targets require the same number of fractions. However, depending on the size of the targets and their proximity to OARs, it might be desirable to treat different targets using different numbers of fractions. The set of targets to be treated would thus vary over the fractions, and consequently, different beams would have to be determined for the different fractions. To adapt the proposed method to this situation, one can apply the method to each unique fraction, considering the set of targets to be treated in each fraction, and possibly a fraction-specific arc trajectory setup.

One implicit limitation of the target partitioning approach is that the same subset of metastases is treated along the entire arc trajectory. 
This might be overly restrictive if island blocking only occurs in a limited number of control points, in which case an alternative idea would be to temporarily close the leaves over selected targets as in intra-arc binary collimation.\cite{macdonald2018intra, lee2022intra} 
Such an approach could potentially allow for eliminating island blocking with fewer arcs, enabling larger average apertures and fewer total MUs, but selecting which targets to cover in which control points is an algorithmic challenge. 
Nevertheless, future studies could explore a combination of the ideas by discounting easily fixable island blocking from the subset costs $c_a(s)$, then initializing the VMAT optimization from intra-arc binary collimation. 
In a similar spirit, the subset costs could be evaluated with respect to optimal dynamic collimator trajectories or optimally tweaked couch angles.

Another implicit limitation of the approach is that each target in $S_a$ is treated exactly once along the associated arc trajectory $a$, i.e., that the subsets in each partition are non-overlapping. This is convenient when all targets require approximately the same number of MUs, making uniform segment weights a reasonable starting guess. However, due to the non-uniform fluence profile of FFF beams, off-center targets may require more MUs than targets near the isocenter. In such cases, it may hypothetically be beneficial to include the off-center targets in more than one subset. 
Future studies may investigate this via manual post-processing, or some variation of the subset optimization strategy detailed in the appendix.

For the simulation experiments, one limitation is that all experiments were run on CT data from a single patient. 
This may, however, be acceptable since we mainly evaluated the dose conformity and dose gradient without considering any particular OAR except the brain. In this context, we expect the size and position of PTVs to be the major source of variability causing different planning challenges across patients.
Although our model for simulated metastases was not systematically validated against clinical data, we note that the range and average of simulated PTV volumes matched well with the clinical case data.
Furthermore, the clinical case study showed no indication that the method would fail to generalize across patients.

One general limitation of our study is that the results can be expected to depend on the VMAT segment conversion algorithm, including the number of fluence optimization iterations before conversion, but we did not evaluate different such algorithms or settings.

Regardless of the potential limitations and future extensions discussed above, 
we anticipate that the target partitioning approach presented in this paper can already be useful in clinical practice.
One major benefit of the proposed algorithm is that it enables some parts of the complex treatment planning process to be decoupled.
In particular, the treatment planner may ignore the island blocking problem while designing the isocenters and arc trajectories, knowing that such issues can later be resolved by the algorithm, and instead focus on aspects such as avoiding beam entrance through important OARs, avoiding collisions, and spreading out the entrance angles sufficiently to allow for conformal dose distributions.

Simplification and automation of the planning process may shorten planning times, freeing valuable resources.
It may also reduce inter-planner variability.
Such improvements facilitate the use of widely available C-arm linacs for multi-target cases, which is essential given the increasing number of patients who can benefit from these treatments.

\section{Conclusions}
Partitioning a set of metastases into smaller subsets to treat with separate arcs can substantially improve the dose gradient compared to simultaneous treatment of all metastases.
The downside is longer treatment times caused by smaller apertures and consequently more MUs.
However, compared to the traditional approach of treating one metastasis at a time, the partitioning approach allows for comparable plan quality with much fewer arcs and MUs.

\subsection*{Acknowledgements}
The authors would like to thank Zacariah Labby and Anna Petoukhova with colleagues for insightful discussions on SRS, and Jakob Ödén for helpful input on radiobiology.

\subsection*{Conflicts of Interest Statement}
Johan Sundström, Anton Finnson, Elin Hynning, and Albin Fredriksson are employed by RaySearch Laboratories AB, Stockholm, Sweden. Anton Finnson owns RaySearch shares. Anton Finnson, Johan Sundström, and Albin Fredriksson are inventors on a pending patent: WO2023213877A1. Albin Fredriksson is an inventor on another related patent: US11607559B2. Geert De Kerf is a member of the RaySearch Laboratories Advisory Board. Geert De Kerf is also involved in a scientific collaboration between Iridium Netwerk and RaySearch Laboratories that includes payments from RaySearch to the institution.

\clearpage

\section*{References}
\addcontentsline{toc}{section}{\numberline{}References}

% The following is when using bibtex and picks up the example.bib file
\bibliography{references.bib}

\begin{thebibliography}{10}

\bibitem{lin2015treatment}
X.~Lin and L.~M. DeAngelis,
\newblock Treatment of brain metastases,
\newblock J. Clin. Oncol. {\bf 33}, 3475 (2015).

\bibitem{lamba2021epidemiology}
N.~Lamba, P.~Y. Wen, and A.~A. Aizer,
\newblock Epidemiology of brain metastases and leptomeningeal disease,
\newblock Neuro-oncology {\bf 23}, 1447--1456 (2021).

\bibitem{gaspar2000validation}
L.~E. Gaspar, C.~Scott, K.~Murray, and W.~Curran,
\newblock Validation of the RTOG recursive partitioning analysis (RPA) classification for brain metastases,
\newblock Int. J. Radiat. Oncol. Biol. Phys. {\bf 47}, 1001--1006 (2000).

\bibitem{sperduto2012summary}
P.~W. Sperduto et~al.,
\newblock Summary report on the graded prognostic assessment: an accurate and facile diagnosis-specific tool to estimate survival for patients with brain metastases,
\newblock J. Clin. Oncol. {\bf 30}, 419 (2012).

\bibitem{suh2020current}
J.~H. Suh, R.~Kotecha, S.~T. Chao, M.~S. Ahluwalia, A.~Sahgal, and E.~L. Chang,
\newblock Current approaches to the management of brain metastases,
\newblock Nature reviews Clinical oncology {\bf 17}, 279--299 (2020).

\bibitem{achrol2019brain}
A.~S. Achrol et~al.,
\newblock Brain metastases,
\newblock Nature Reviews Disease Primers {\bf 5}, 5 (2019).

\bibitem{vogelbaum2022treatment}
M.~A. Vogelbaum et~al.,
\newblock Treatment for brain metastases: ASCO-SNO-ASTRO guideline, 2022.

\bibitem{yamamoto2014stereotactic}
M.~Yamamoto et~al.,
\newblock Stereotactic radiosurgery for patients with multiple brain metastases (JLGK0901): a multi-institutional prospective observational study,
\newblock The Lancet Oncology {\bf 15}, 387--395 (2014).

\bibitem{leksell1951stereotaxic}
L.~Leksell,
\newblock The stereotaxic method and radiosurgery of the brain,
\newblock Acta chir scand {\bf 102}, 316--319 (1951).

\bibitem{ICRU_91}
ICRU,
\newblock ICRU Report 91: Prescribing, Recording, and Reporting of Stereotactic Treatments with Small Photon Beams,
\newblock Journal of the ICRU {\bf 14}, 1--145 (2014).

\bibitem{gill1991relocatable}
S.~Gill, D.~Thomas, A.~Warrington, and M.~Brada,
\newblock Relocatable frame for stereotactic external beam radiotherapy,
\newblock Int. J. Radiat. Oncol. Biol. Phys. {\bf 20}, 599--603 (1991).

\bibitem{fiagbedzi2023radiotherapy}
E.~Fiagbedzi, F.~Hasford, S.~N. Tagoe, and A.~Nisbet,
\newblock Radiotherapy infrastructure for brain metastasis treatment in Africa: practical guildelines for implementation of a stereotactic radiosurgery (SRS) program,
\newblock Health and Technology {\bf 13}, 893--904 (2023).

\bibitem{clark2010feasibility}
G.~M. Clark, R.~A. Popple, P.~E. Young, and J.~B. Fiveash,
\newblock Feasibility of single-isocenter volumetric modulated arc radiosurgery for treatment of multiple brain metastases,
\newblock Int. J. Radiat. Oncol. Biol. Phys. {\bf 76}, 296--302 (2010).

\bibitem{vergalasova2019multi}
I.~Vergalasova et~al.,
\newblock Multi-institutional dosimetric evaluation of modern day stereotactic radiosurgery (SRS) treatment options for multiple brain metastases,
\newblock Frontiers in oncology {\bf 9}, 483 (2019).

\bibitem{huang2014radiosurgery}
Y.~Huang, K.~Chin, J.~R. Robbins, J.~Kim, H.~Li, H.~Amro, I.~J. Chetty, J.~Gordon, and S.~Ryu,
\newblock Radiosurgery of multiple brain metastases with single-isocenter dynamic conformal arcs (SIDCA),
\newblock Radiother. Oncol. {\bf 112}, 128--132 (2014).

\bibitem{gevaert2016evaluation}
T.~Gevaert, F.~Steenbeke, L.~Pellegri, B.~Engels, N.~Christian, M.-T. Hoornaert, D.~Verellen, C.~Mitine, and M.~De~Ridder,
\newblock Evaluation of a dedicated brain metastases treatment planning optimization for radiosurgery: a new treatment paradigm?,
\newblock Radiation Oncology {\bf 11}, 1--7 (2016).

\bibitem{kang2010}
J.~Kang, E.~C. Ford, K.~Smith, J.~Wong, and T.~R. McNutt,
\newblock A method for optimizing LINAC treatment geometry for volumetric modulated arc therapy of multiple brain metastases,
\newblock Med. Phys. {\bf 37}, 4146--4154 (2010).

\bibitem{ohira2018hyperarc}
S.~Ohira, Y.~Ueda, Y.~Akino, M.~Hashimoto, A.~Masaoka, T.~Hirata, M.~Miyazaki, M.~Koizumi, and T.~Teshima,
\newblock HyperArc VMAT planning for single and multiple brain metastases stereotactic radiosurgery: a new treatment planning approach,
\newblock Radiation Oncology {\bf 13}, 1--9 (2018).

\bibitem{wu2016optimization}
Q.~Wu, K.~C. Snyder, C.~Liu, Y.~Huang, B.~Zhao, I.~J. Chetty, and N.~Wen,
\newblock Optimization of treatment geometry to reduce normal brain dose in Radiosurgery of multiple brain metastases with single--Isocenter volumetric modulated arc therapy,
\newblock Scientific reports {\bf 6}, 34511 (2016).

\bibitem{macdonald2018dynamic}
R.~L. MacDonald, C.~G. Thomas, and A.~Syme,
\newblock Dynamic collimator trajectory algorithm for multiple metastases dynamic conformal arc treatment planning,
\newblock Med. Phys. {\bf 45}, 5--17 (2018).

\bibitem{battinelli2021collimator}
C.~Battinelli, A.~Fredriksson, and K.~Eriksson,
\newblock Collimator angle optimization for multiple brain metastases in dynamic conformal arc treatment planning,
\newblock Med. Phys. {\bf 48}, 5414--5422 (2021).

\bibitem{chang2018restricted}
J.~Chang, A.~G. Wernicke, and S.~C. Pannullo,
\newblock Restricted single isocenter for multiple targets dynamic conformal arc (RSIMT DCA) technique for brain stereotactic radiosurgery (SRS) planning,
\newblock Journal of Radiosurgery and SBRT {\bf 5}, 145 (2018).

\bibitem{palmiero2021management}
A.~N. Palmiero, D.~Fabian, W.~St~Clair, M.~Randall, and D.~Pokhrel,
\newblock Management of multiple brain metastases via dual-isocenter VMAT stereotactic radiosurgery,
\newblock Med. Dosim {\bf 46}, 240--246 (2021).

\bibitem{yun1986dynamic}
D.~Yun~Yeh,
\newblock A dynamic programming approach to the complete set partitioning problem,
\newblock BIT Numerical Mathematics {\bf 26}, 467--474 (1986).

\bibitem{michalak2016hybrid}
T.~Michalak, T.~Rahwan, E.~Elkind, M.~Wooldridge, and N.~R. Jennings,
\newblock A hybrid exact algorithm for complete set partitioning,
\newblock Artificial Intelligence {\bf 230}, 14--50 (2016).

\bibitem{buller2020brain}
M.~Buller, K.~Chapple, and C.~Bird,
\newblock Brain Metastases: Insights from Statistical Modeling of Size Distribution,
\newblock American Journal of Neuroradiology {\bf 41}, 579--582 (2020).

\bibitem{gondi2022ASTROguidelines}
V.~Gondi et~al.,
\newblock Radiation therapy for brain metastases: an ASTRO clinical practice guideline,
\newblock Practical radiation oncology {\bf 12}, 265--282 (2022).

\bibitem{papenberg2021anticlustering}
M.~Papenberg and G.~W. Klau,
\newblock Using anticlustering to partition data sets into equivalent parts,
\newblock Psychological Methods {\bf 26}, 161--174 (2021).

\bibitem{morrison2016single}
J.~Morrison, R.~Hood, F.-F. Yin, J.~K. Salama, J.~Kirkpatrick, and J.~Adamson,
\newblock Is a single isocenter sufficient for volumetric modulated arc therapy radiosurgery when multiple intracranial metastases are spatially dispersed?,
\newblock Med. Dosim {\bf 41}, 285--289 (2016).

\bibitem{ballangrud2018institutional}
{\AA}.~Ballangrud, L.~C. Kuo, L.~Happersett, S.~B. Lim, K.~Beal, Y.~Yamada, M.~Hunt, and J.~Mechalakos,
\newblock Institutional experience with SRS VMAT planning for multiple cranial metastases,
\newblock J of App Clin Med Phys {\bf 19}, 176--183 (2018).

\bibitem{paddick2000conformityIndex}
I.~Paddick,
\newblock A simple scoring ratio to index the conformity of radiosurgical treatment plans,
\newblock Journal of neurosurgery {\bf 93}, 219--222 (2000).

\bibitem{paddick2006gradientIndex}
I.~Paddick and B.~Lippitz,
\newblock A simple dose gradient measurement tool to complement the conformity index,
\newblock Journal of neurosurgery {\bf 105}, 194--201 (2006).

\bibitem{cui2020retrospective}
Y.~Cui, H.~Gao, J.~Zhang, J.~P. Kirkpatrick, and F.-F. Yin,
\newblock Retrospective quality metrics review of stereotactic radiosurgery plans treating multiple targets using single-isocenter volumetric modulated arc therapy,
\newblock J of App Clin Med Phys {\bf 21}, 93--99 (2020).

\bibitem{desai2022estimate}
D.~D. Desai and I.~L. Cordrey,
\newblock How to estimate R50\% for cranial SRS/SRT cases with overlapping 50\% isodose volumes: A proposed system,
\newblock J of App Clin Med Phys {\bf 23}, e13624 (2022).

\bibitem{dimitriadis2018novel}
A.~Dimitriadis and I.~Paddick,
\newblock A novel index for assessing treatment plan quality in stereotactic radiosurgery,
\newblock Journal of neurosurgery {\bf 129}, 118--124 (2018).

\bibitem{holm1979simple}
S.~Holm,
\newblock A simple sequentially rejective multiple test procedure,
\newblock Scandinavian journal of statistics , 65--70 (1979).

\bibitem{blonigen2010irradiated}
B.~J. Blonigen, R.~D. Steinmetz, L.~Levin, M.~A. Lamba, R.~E. Warnick, and J.~C. Breneman,
\newblock Irradiated volume as a predictor of brain radionecrosis after linear accelerator stereotactic radiosurgery,
\newblock Int. J. Radiat. Oncol. Biol. Phys. {\bf 77}, 996--1001 (2010).

\bibitem{milano2021single}
M.~T. Milano et~al.,
\newblock Single-and multifraction stereotactic radiosurgery dose/volume tolerances of the brain,
\newblock Int. J. Radiat. Oncol. Biol. Phys. {\bf 110}, 68--86 (2021).

\bibitem{unkelbach2015optimization}
J.~Unkelbach et~al.,
\newblock Optimization approaches to volumetric modulated arc therapy planning,
\newblock Med. Phys. {\bf 42}, 1367--1377 (2015).

\bibitem{fowler2004loss}
J.~F. Fowler, J.~S. Welsh, and S.~P. Howard,
\newblock Loss of biological effect in prolonged fraction delivery,
\newblock Int. J. Radiat. Oncol. Biol. Phys. {\bf 59}, 242--249 (2004).

\bibitem{nakano2022radiobiological}
H.~Nakano et~al.,
\newblock Radiobiological evaluation considering the treatment time with stereotactic radiosurgery for brain metastases,
\newblock BJR| Open {\bf 4}, 20220013 (2022).

\bibitem{ianiro2023multiple}
A.~Ianiro, E.~Infusino, M.~D’Andrea, L.~Marucci, A.~Farneti, F.~Sperati, B.~Cassano, S.~Ungania, and A.~Soriani,
\newblock Multiple brain metastases radiosurgery with cyberKnife device: dosimetric comparison between fixed/iris and multileaf collimator plans,
\newblock Journal of Medical Physics {\bf 48}, 120--128 (2023).

\bibitem{paddick2023benchmarking}
I.~Paddick, J.~Mott, J.~Bedford, P.~Filatov, D.~Grishchuk, G.~Orchin, P.~Houston, and D.~J. Eaton,
\newblock Benchmarking tests of contemporary SRS platforms: have technological developments resulted in improved treatment plan quality?,
\newblock Practical Radiation Oncology {\bf 13}, e451--e459 (2023).

\bibitem{niu2024comparative}
Y.~Niu, A.~Rashid, J.-m. Lee, M.~Carrasquilla, D.~R. Conroy, B.~T. Collins, A.~Satinsky, K.~R. Unger, and D.~Pang,
\newblock Comparative analysis of plan quality and delivery efficiency: ZAP-X vs. CyberKnife for brain metastases treatment,
\newblock Frontiers in Oncology {\bf 14}, 1333642 (2024).

\bibitem{macdonald2018intra}
R.~L. MacDonald, C.~G. Thomas, L.~Ward, and A.~Syme,
\newblock Intra-arc binary collimation algorithm for the optimization of stereotactic radiotherapy treatment of multiple metastases with multiple prescriptions,
\newblock Med. Phys. {\bf 45}, 5597--5607 (2018).

\bibitem{lee2022intra}
E.~Lee, R.~L. MacDonald, C.~G. Thomas, and A.~Syme,
\newblock Intra-arc binary collimation with dynamic axes trajectory optimization for the SRS treatment of multiple metastases with multiple prescriptions,
\newblock Med. Phys. {\bf 49}, 4305--4321 (2022).

\end{thebibliography}
\bibliographystyle{medphy.bst}

\begin{customappendicesenvironment}{Appendix}{\sf \Alph{subsection}}
\subsection{Generalized algorithm for multiple arc trajectories}
\label{appendix:multiple_trajectories}
The multi-target partitioning algorithm is here extended to handle a set $A$ of arc trajectories. 
We define a cost function $f_a$ as in the main text for each trajectory $a\in A$ and sum these to create a composite objective.
Such an objective is separable and can be minimized by running the single-arc algorithm once per trajectory, given user defined $S_a$ and $n_a$ as input.

The choice of $S_a$ and $n_a$ leaves some control to the planner.
For example, as explored in our experiments, one may try all combinations of $n_a$ summing to some value $N$ representing the total number of arcs.
Another possibility is to design four arc trajectories but then include each metastasis in just three of the associated sets $S_a$.
Instead of manually exploring all such configurations, we let $S_a$ and $n_a$ be optimization variables in a generalized problem:
\begin{equation}
\label{eq:partitioning_multiple_arcs}
\begin{aligned}
    \displaystyle \min_{((S_a, n_a))_{a\in A}} \quad & \sum_{a\in A} f_a(S_a, n_a)
    & \\
    \textrm{s.t.}
    \quad & \setofpartitions{S_a}{n_a} \neq \varnothing, &\forall a\in A \\
    \quad & S_a^{\min} \subseteq S_a \subseteq S_a^{\max}, \quad &\forall a\in A \\
    \quad & n_a^{\min} \leq n_a \leq n_a^{\max}, \quad &\forall a\in A \\
    \quad & N^{\min} \leq \textstyle \sum_a n_a \leq N^{\max} & 
\end{aligned}
\end{equation}
where the bounds $S_a^{\min}$, $S_a^{\max}$, $n_a^{\min}$, $n_a^{\max}$, $N^{\min}$, and $N^{\max}$ are specified by the user.
Note that the total number of arcs $\textstyle \sum_a n_a$ can be penalized by adding a constant to the subset cost functions $c_a$ in the definition of $f_a$.

To solve the generalized problem \eqref{eq:partitioning_multiple_arcs}, we first evaluate
$f_a\left(S_a^{\max},\, \min\left(\abs{S_a^{\max}}, n_a^{\max}\right)\right)$
for all $a\in A$ using our recursive approach.
This effectively computes $f_a(S_a, n_a)$ for all interesting combinations of $S_a$ and $n_a$, and by storing these results in memory, we enable computationally efficient evaluations of the composite objective function.
As long as the domain is reasonably small, this allows the generalized problem to be solved to optimality in reasonable time by exhaustively searching through the domain. Such a brute-force approach was used in our study, where the domain was limited by $S_a^{\min} = S_a^{\max}$ and $N^{\min} = N^{\max}$.

One interesting special case is $S_a = \varnothing$, which corresponds to discarding arc trajectory~$a$.
For example, this could be used to select 3 out of 4 candidate trajectories.
Note that $\setofpartitions{\varnothing}{0} = \{\varnothing \} \neq \varnothing$ and that $f_a(\varnothing, 0) = 0$. 
For such use cases, it may be useful to include additional constraints, e.g., that each target must be in a minimum number of subsets. 

Another interesting use case is fractionated treatments with non-identical fractions.
This may be modeled in \eqref{eq:partitioning_multiple_arcs} by letting $A$ include separate (but possibly geometrically identical) arc trajectories for each fraction, using $S_a^{\min}$ and $S_a^{\max}$ to limit which targets may and must be treated in each fraction.
For example, one can exclude all targets that do not require the full number of fractions from all $S_a^{\min}$ and instead enforce additional constraints on the allowed combinations of $S_a$, in particular, that each metastasis should be treated by sufficiently many arcs in exactly the right number of fractions.
Note that compared to setting all $S_a^{\min} = S_a^{\max}$, this does not affect the precomputation of $f_a(S_a, n_a)$ in any way. Only the subsequent exhaustive search is affected.

\end{customappendicesenvironment}

\ifpreprintelse{
\clearpage
\pagestyle{fancy}
\pagenumbering{arabic} % reset page number
\renewcommand*{\thepage}{\sf S.\arabic{page}}

%%%%%%%%%%%%%%%%%%%%%%%%%%%%%%%%%%%%%%%%%%%%%%%%%%%%%%%%%%%%%%%%%%%%%%%%%%%%%%%
% Headers and footers
%%%%%%%%%%%%%%%%%%%%%%%%%%%%%%%%%%%%%%%%%%%%%%%%%%%%%%%%%%%%%%%%%%%%%%%%%%%%%%%

\lhead[{\sffamily page~\thepage}]{{\sffamily \itshape Partitioning of multiple brain metastases: supplementary material}}

%%%%%%%%%%%%%%%%%%%%%%%%%%%%%%%%%%%%%%%%%%%%%%%%%%%%%%%%%%%%%%%%%%%%%%%%%%%%%%%
% Table definition
%%%%%%%%%%%%%%%%%%%%%%%%%%%%%%%%%%%%%%%%%%%%%%%%%%%%%%%%%%%%%%%%%%%%%%%%%%%%%%%
\newcommand{\statstable}[3]{
\vspace*{2ex}
\begin{table}[!htbp]
\centering
\captionv{16}{}
{#1} % long title
#3 % \label{...}
\vspace*{2ex}
\renewrobustcmd{\bfseries}{\fontseries{b}\selectfont}
\sisetup{
    mode=text,
    group-digits=integer, % https://tex.stackexchange.com/a/561368
    group-minimum-digits=5,
    tight-spacing=true,
    round-mode=places,
    detect-weight=true,
}
\footnotesize
#2 % tabular
\end{table}
}

%%%%%%%%%%%%%%%%%%%%%%%%%%%%%%%%%%%%%%%%%%%%%%%%%%%%%%%%%%%%%%%%%%%%%%%%%%%%%%%
%%%%%%%%%%%%%%%%%%%%%%%%%%%%%%%%%%%%%%%%%%%%%%%%%%%%%%%%%%%%%%%%%%%%%%%%%%%%%%%
\begin{customappendicesenvironment}{Supplementary material}{{\sf S.\Roman{subsection}}}
\subsection{Statistical analysis}
\label{appendix:statistical_analysis}
In this section, we show the statistical analysis of the results from the simulation experiments.
The distributions of pairwise differences between the baselines and the proposed algorithm, per metric and per number of arcs, are illustrated in Figure~\ref{fig:diff_ptv_metrics} and \ref{fig:diff_plan_metrics}.

\begin{figure}[h]
% \vspace*{2ex}
\begin{center}
\includegraphics[width=\linewidth]{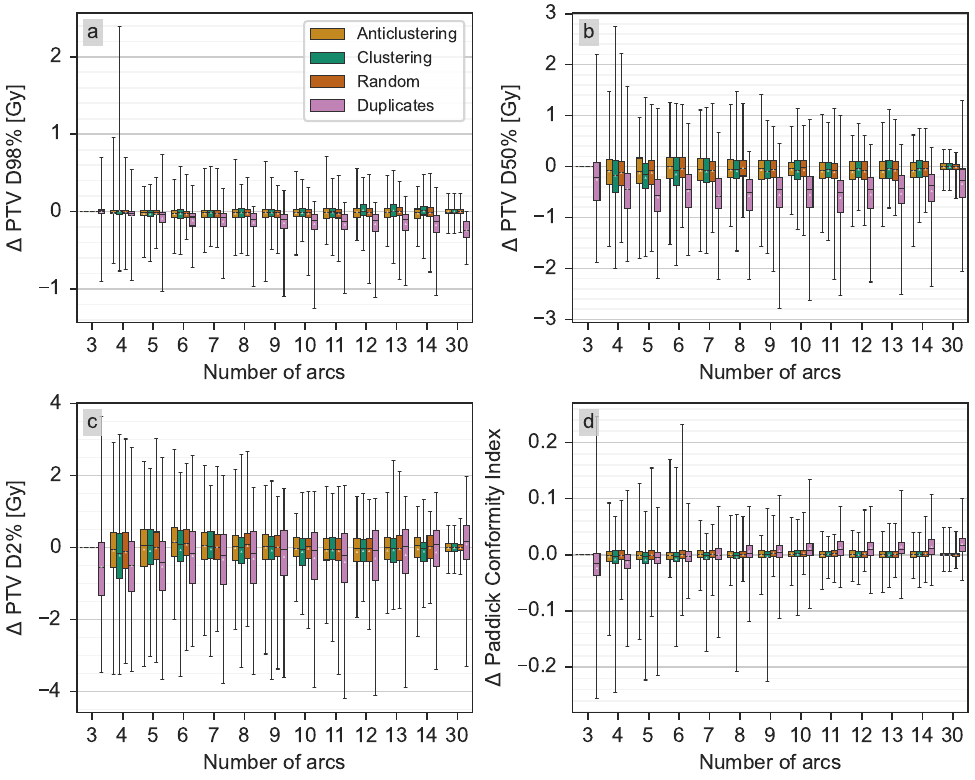}
\captionv{16}{}
{
    \label{fig:diff_ptv_metrics}
    Distributions of pairwise differences between the baseline algorithms and the proposed algorithm ($\text{baseline}-\text{proposed}$) for the PTV-specific metrics. Each individual box plot is based on $20\times10=200$ data points. 
% \vspace*{2ex}
}
\end{center}
\end{figure}

\begin{figure}[ht]
% \vspace*{2ex}
\begin{center}
\includegraphics[width=\linewidth]{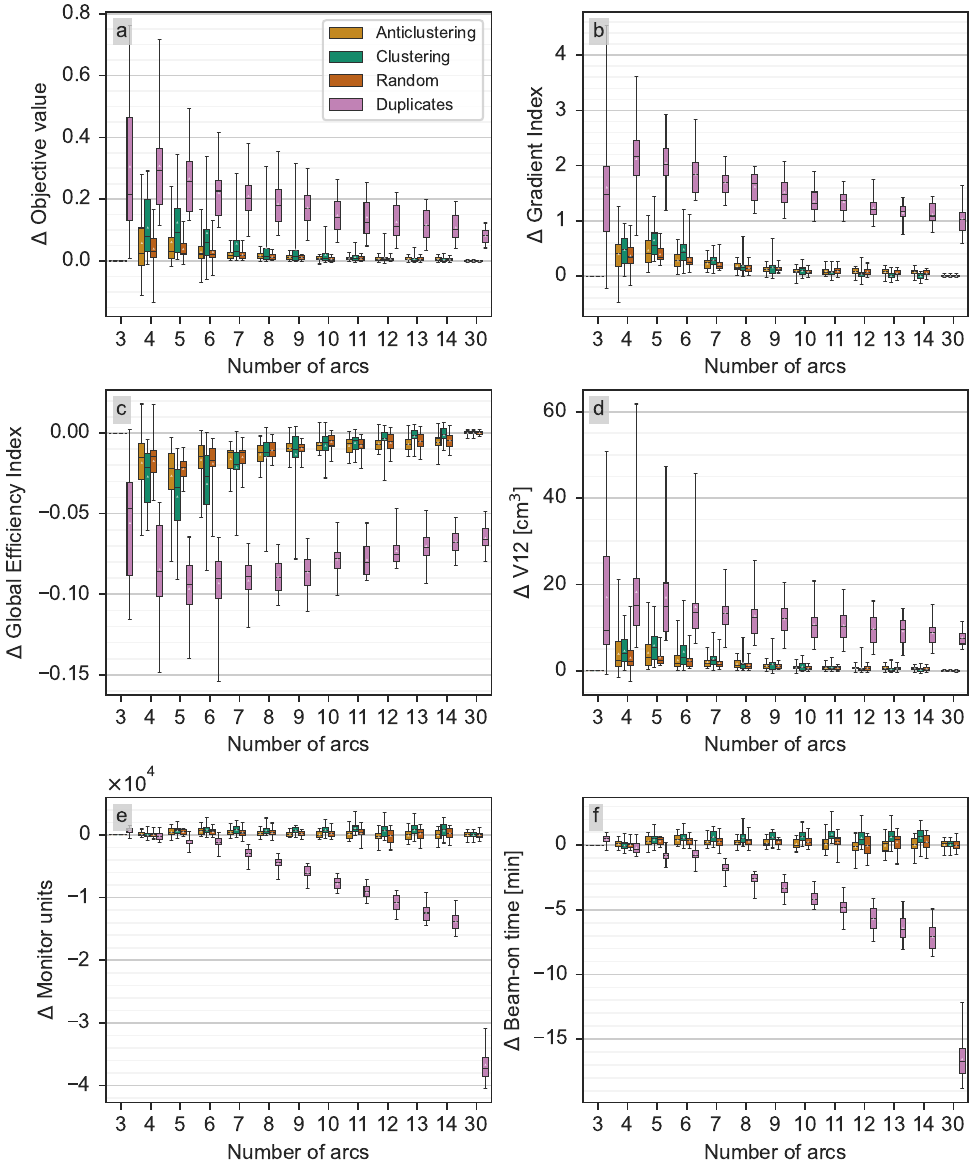}
\captionv{16}{}
{
    \label{fig:diff_plan_metrics}
    Distribution of pairwise differences between the baseline algorithms and the proposed algorithm ($\text{baseline}-\text{proposed}$) for the plan-specific metrics. Each individual box plot is based on $20$ data points.
% \vspace*{2ex}
}
\end{center}
\end{figure}

\FloatBarrier
The median pairwise differences between the baselines and the proposed algorithm are repeated in Tables~\ref{tab:stats_ptv_d98}, \ref{tab:stats_ptv_d50}, \ref{tab:stats_ptv_d2}, \ref{tab:stats_paddick_conformity_index}, \ref{tab:stats_objective_value}, \ref{tab:stats_gradient_index}, \ref{tab:stats_global_efficiency_index}, \ref{tab:stats_v12}, \ref{tab:stats_monitor_units}, and \ref{tab:stats_beamon_time} together with the associated $p$-values. 
The $p$-values are from the Wilcoxon signed-rank tests with null hypothesis that the distribution of pairwise differences is symmetric around zero.
Statistically significant results are shown in bold.
As discussed in the main text, each individual significance test used a corrected $p$-value in the range $[\alpha/m, \alpha]$, where $m$ is the number of table rows, such that the probability of one or more false positives is limited to $\alpha=0.05$ per column. Significant results are not interesting if the effect is small. The magnitude of the median pairwise differences can for example be put in relation to the mean absolute values for the proposed algorithm presented in the main text.

\statstable
{Median pairwise differences in PTV $\text{D}_{98\%}$ between the baseline algorithms and the proposed algorithm. Statistically significant results are shown in bold.}
{\begin{tabular}{S[table-format=2.0, round-precision=0]
|S[table-format=-2.2, round-precision=2]S[table-format=1.4, round-precision=4]
|S[table-format=-2.2, round-precision=2]S[table-format=1.4, round-precision=4]
|S[table-format=-2.2, round-precision=2]S[table-format=1.4, round-precision=4]
|S[table-format=-2.2, round-precision=2]S[table-format=1.4, round-precision=4]}
\toprule
{} & \multicolumn{2}{c}{Anticlustering} & \multicolumn{2}{c}{Clustering} & \multicolumn{2}{c}{Random} & \multicolumn{2}{c}{Duplicates} \\
{Arcs} & {$\Delta$$\text{D}_{98\%}$} & {$p$-value} & {$\Delta$$\text{D}_{98\%}$} & {$p$-value} & {$\Delta$$\text{D}_{98\%}$} & {$p$-value} & {$\Delta$$\text{D}_{98\%}$} & {$p$-value} \\
\midrule
3 &  &  &  &  &  &  & 0.002178 & 0.888174 \\
4 & \bfseries -0.007893 & \bfseries 0.004693 & -0.004469 & 0.104592 & -0.006294 & 0.065239 & \bfseries -0.018435 & \bfseries 0.000000 \\
5 & -0.006249 & 0.012349 & \bfseries -0.010531 & \bfseries 0.000148 & \bfseries -0.008427 & \bfseries 0.000301 & \bfseries -0.039340 & \bfseries 0.000000 \\
6 & \bfseries -0.024130 & \bfseries 0.000013 & \bfseries -0.013455 & \bfseries 0.003968 & \bfseries -0.023623 & \bfseries 0.000000 & \bfseries -0.069679 & \bfseries 0.000000 \\
7 & \bfseries -0.013569 & \bfseries 0.000227 & \bfseries -0.015750 & \bfseries 0.000444 & \bfseries -0.019246 & \bfseries 0.000036 & \bfseries -0.072468 & \bfseries 0.000000 \\
8 & \bfseries -0.017107 & \bfseries 0.001582 & -0.010787 & 0.127247 & \bfseries -0.014310 & \bfseries 0.003551 & \bfseries -0.096526 & \bfseries 0.000000 \\
9 & -0.006594 & 0.067783 & -0.006486 & 0.096230 & \bfseries -0.012886 & \bfseries 0.001397 & \bfseries -0.100292 & \bfseries 0.000000 \\
10 & -0.015160 & 0.007695 & -0.015039 & 0.030349 & \bfseries -0.012577 & \bfseries 0.002368 & \bfseries -0.111808 & \bfseries 0.000000 \\
11 & -0.011636 & 0.024973 & -0.005210 & 0.191089 & \bfseries -0.017770 & \bfseries 0.001104 & \bfseries -0.130792 & \bfseries 0.000000 \\
12 & -0.010990 & 0.082833 & 0.003149 & 0.157803 & -0.007482 & 0.102772 & \bfseries -0.111313 & \bfseries 0.000000 \\
13 & -0.008143 & 0.212218 & 0.001204 & 0.291068 & -0.000747 & 0.901655 & \bfseries -0.105106 & \bfseries 0.000000 \\
14 & -0.014801 & 0.009133 & 0.011064 & 0.204269 & -0.007761 & 0.209100 & \bfseries -0.123386 & \bfseries 0.000000 \\
30 &  &  &  &  &  &  & \bfseries -0.241656 & \bfseries 0.000000 \\
\bottomrule
{-} & {\unit{\gray}} & {-} & {\unit{\gray}} & {-} & {\unit{\gray}} & {-} & {\unit{\gray}} & {-}
\end{tabular}
}
{\label{tab:stats_ptv_d98}}

\statstable
{Median pairwise differences in PTV $\text{D}_{50\%}$ between the baseline algorithms and the proposed algorithm. Statistically significant results are shown in bold.}
{\begin{tabular}{S[table-format=2.0, round-precision=0]
|S[table-format=-2.2, round-precision=2]S[table-format=1.4, round-precision=4]
|S[table-format=-2.2, round-precision=2]S[table-format=1.4, round-precision=4]
|S[table-format=-2.2, round-precision=2]S[table-format=1.4, round-precision=4]
|S[table-format=-2.2, round-precision=2]S[table-format=1.4, round-precision=4]}
\toprule
{} & \multicolumn{2}{c}{Anticlustering} & \multicolumn{2}{c}{Clustering} & \multicolumn{2}{c}{Random} & \multicolumn{2}{c}{Duplicates} \\
{Arcs} & {$\Delta$$\text{D}_{50\%}$} & {$p$-value} & {$\Delta$$\text{D}_{50\%}$} & {$p$-value} & {$\Delta$$\text{D}_{50\%}$} & {$p$-value} & {$\Delta$$\text{D}_{50\%}$} & {$p$-value} \\
\midrule
3 &  &  &  &  &  &  & \bfseries -0.215883 & \bfseries 0.000000 \\
4 & \bfseries -0.085308 & \bfseries 0.004514 & \bfseries -0.179845 & \bfseries 0.000001 & \bfseries -0.117465 & \bfseries 0.000016 & \bfseries -0.457330 & \bfseries 0.000000 \\
5 & \bfseries -0.093240 & \bfseries 0.000423 & \bfseries -0.148462 & \bfseries 0.000000 & \bfseries -0.073730 & \bfseries 0.000284 & \bfseries -0.572683 & \bfseries 0.000000 \\
6 & 0.000557 & 0.762232 & \bfseries -0.075543 & \bfseries 0.001006 & -0.033018 & 0.113797 & \bfseries -0.458420 & \bfseries 0.000000 \\
7 & \bfseries -0.058998 & \bfseries 0.002207 & \bfseries -0.072379 & \bfseries 0.002091 & \bfseries -0.078973 & \bfseries 0.000266 & \bfseries -0.595366 & \bfseries 0.000000 \\
8 & \bfseries -0.062958 & \bfseries 0.004192 & \bfseries -0.053098 & \bfseries 0.003580 & -0.030023 & 0.166993 & \bfseries -0.501558 & \bfseries 0.000000 \\
9 & \bfseries -0.043127 & \bfseries 0.004585 & \bfseries -0.087788 & \bfseries 0.001542 & \bfseries -0.053304 & \bfseries 0.004973 & \bfseries -0.450544 & \bfseries 0.000000 \\
10 & \bfseries -0.038929 & \bfseries 0.001253 & \bfseries -0.069891 & \bfseries 0.000024 & \bfseries -0.021132 & \bfseries 0.015763 & \bfseries -0.454635 & \bfseries 0.000000 \\
11 & \bfseries -0.075831 & \bfseries 0.000012 & \bfseries -0.066180 & \bfseries 0.000123 & \bfseries -0.071989 & \bfseries 0.000001 & \bfseries -0.517224 & \bfseries 0.000000 \\
12 & \bfseries -0.082328 & \bfseries 0.000014 & \bfseries -0.029291 & \bfseries 0.010354 & \bfseries -0.072168 & \bfseries 0.000043 & \bfseries -0.442444 & \bfseries 0.000000 \\
13 & \bfseries -0.074761 & \bfseries 0.000021 & \bfseries -0.039587 & \bfseries 0.002801 & \bfseries -0.049082 & \bfseries 0.001752 & \bfseries -0.435142 & \bfseries 0.000000 \\
14 & \bfseries -0.070624 & \bfseries 0.000218 & \bfseries -0.044159 & \bfseries 0.002801 & \bfseries -0.038367 & \bfseries 0.005030 & \bfseries -0.378164 & \bfseries 0.000000 \\
30 &  &  &  &  &  &  & \bfseries -0.282390 & \bfseries 0.000000 \\
\bottomrule
{-} & {\unit{\gray}} & {-} & {\unit{\gray}} & {-} & {\unit{\gray}} & {-} & {\unit{\gray}} & {-}
\end{tabular}
}
{\label{tab:stats_ptv_d50}}

\statstable
{Median pairwise differences in PTV $\text{D}_{2\%}$ between the baseline algorithms and the proposed algorithm. Statistically significant results are shown in bold.}
{\begin{tabular}{S[table-format=2.0, round-precision=0]
|S[table-format=-2.2, round-precision=2]S[table-format=1.4, round-precision=4]
|S[table-format=-2.2, round-precision=2]S[table-format=1.4, round-precision=4]
|S[table-format=-2.2, round-precision=2]S[table-format=1.4, round-precision=4]
|S[table-format=-2.2, round-precision=2]S[table-format=1.4, round-precision=4]}
\toprule
{} & \multicolumn{2}{c}{Anticlustering} & \multicolumn{2}{c}{Clustering} & \multicolumn{2}{c}{Random} & \multicolumn{2}{c}{Duplicates} \\
{Arcs} & {$\Delta$$\text{D}_{2\%}$} & {$p$-value} & {$\Delta$$\text{D}_{2\%}$} & {$p$-value} & {$\Delta$$\text{D}_{2\%}$} & {$p$-value} & {$\Delta$$\text{D}_{2\%}$} & {$p$-value} \\
\midrule
3 &  &  &  &  &  &  & \bfseries -0.551506 & \bfseries 0.000000 \\
4 & -0.058774 & 0.344166 & \bfseries -0.168695 & \bfseries 0.003099 & -0.099069 & 0.027810 & \bfseries -0.495559 & \bfseries 0.000000 \\
5 & 0.067596 & 0.864183 & 0.045706 & 0.950977 & 0.002964 & 0.709920 & \bfseries -0.403987 & \bfseries 0.000000 \\
6 & \bfseries 0.145287 & \bfseries 0.000266 & 0.089265 & 0.508824 & 0.112014 & 0.007472 & \bfseries -0.153348 & \bfseries 0.001338 \\
7 & 0.051982 & 0.146262 & 0.040881 & 0.650144 & 0.012699 & 0.719919 & \bfseries -0.213826 & \bfseries 0.000052 \\
8 & 0.025210 & 0.917097 & -0.009384 & 0.235940 & 0.088392 & 0.214014 & \bfseries -0.176165 & \bfseries 0.000141 \\
9 & 0.040400 & 0.516672 & 0.023704 & 0.906477 & 0.005424 & 0.434604 & -0.061410 & 0.018637 \\
10 & -0.022722 & 0.786474 & -0.049365 & 0.012787 & 0.016189 & 0.997572 & \bfseries -0.093579 & \bfseries 0.006463 \\
11 & -0.057449 & 0.081109 & -0.042183 & 0.121549 & -0.056191 & 0.063645 & \bfseries -0.227618 & \bfseries 0.000019 \\
12 & -0.034075 & 0.052909 & -0.029189 & 0.063295 & -0.025995 & 0.122139 & \bfseries -0.074677 & \bfseries 0.000731 \\
13 & -0.000366 & 0.201237 & -0.040480 & 0.026863 & -0.026517 & 0.428180 & 0.031257 & 0.250234 \\
14 & 0.043159 & 0.793975 & -0.039216 & 0.011884 & 0.024353 & 0.683802 & 0.074785 & 0.800555 \\
30 &  &  &  &  &  &  & \bfseries 0.157582 & \bfseries 0.009744 \\
\bottomrule
{-} & {\unit{\gray}} & {-} & {\unit{\gray}} & {-} & {\unit{\gray}} & {-} & {\unit{\gray}} & {-}
\end{tabular}
}
{\label{tab:stats_ptv_d2}}

\statstable
{Median pairwise differences in Paddick Conformity Index between the baseline algorithms and the proposed algorithm. Statistically significant results are shown in bold.}
{\begin{tabular}{S[table-format=2.0, round-precision=0]
|S[table-format=-1.3, round-precision=3]S[table-format=1.4, round-precision=4]
|S[table-format=-1.3, round-precision=3]S[table-format=1.4, round-precision=4]
|S[table-format=-1.3, round-precision=3]S[table-format=1.4, round-precision=4]
|S[table-format=-1.3, round-precision=3]S[table-format=1.4, round-precision=4]}
\toprule
{} & \multicolumn{2}{c}{Anticlustering} & \multicolumn{2}{c}{Clustering} & \multicolumn{2}{c}{Random} & \multicolumn{2}{c}{Duplicates} \\
{Arcs} & {$\Delta$PCI} & {$p$-value} & {$\Delta$PCI} & {$p$-value} & {$\Delta$PCI} & {$p$-value} & {$\Delta$PCI} & {$p$-value} \\
\midrule
3 &  &  &  &  &  &  & \bfseries -0.014842 & \bfseries 0.000000 \\
4 & -0.001497 & 0.080683 & \bfseries -0.002637 & \bfseries 0.005030 & -0.001201 & 0.659818 & \bfseries -0.009673 & \bfseries 0.000000 \\
5 & -0.000862 & 0.162533 & \bfseries -0.002848 & \bfseries 0.000182 & -0.000995 & 0.616325 & \bfseries -0.005515 & \bfseries 0.000102 \\
6 & -0.001660 & 0.021256 & \bfseries -0.003441 & \bfseries 0.000097 & 0.000094 & 0.984945 & \bfseries -0.003471 & \bfseries 0.005663 \\
7 & 0.001005 & 0.239337 & -0.001715 & 0.009068 & -0.000221 & 0.349166 & -0.000693 & 0.684697 \\
8 & -0.000433 & 0.600979 & -0.000816 & 0.225020 & 0.000299 & 0.916131 & \bfseries 0.003129 & \bfseries 0.000096 \\
9 & -0.000297 & 0.428180 & 0.000483 & 0.762232 & \bfseries 0.001641 & \bfseries 0.001915 & \bfseries 0.003536 & \bfseries 0.000006 \\
10 & 0.001284 & 0.008713 & \bfseries 0.001974 & \bfseries 0.003440 & \bfseries 0.001057 & \bfseries 0.004603 & \bfseries 0.006939 & \bfseries 0.000000 \\
11 & 0.000366 & 0.556827 & 0.000905 & 0.051864 & 0.001504 & 0.006560 & \bfseries 0.009812 & \bfseries 0.000000 \\
12 & 0.000069 & 0.360585 & -0.000432 & 0.978147 & 0.000736 & 0.073510 & \bfseries 0.008780 & \bfseries 0.000000 \\
13 & 0.000408 & 0.581626 & -0.000609 & 0.996600 & 0.000570 & 0.093325 & \bfseries 0.010089 & \bfseries 0.000000 \\
14 & 0.000558 & 0.350424 & -0.000121 & 0.496397 & 0.000949 & 0.019137 & \bfseries 0.011646 & \bfseries 0.000000 \\
30 &  &  &  &  &  &  & \bfseries 0.017018 & \bfseries 0.000000 \\
\bottomrule
{-} & {-} & {-} & {-} & {-} & {-} & {-} & {-} & {-}
\end{tabular}
}
{\label{tab:stats_paddick_conformity_index}}

\statstable
{Median pairwise differences in objective value between the baseline algorithms and the proposed algorithm. Statistically significant results are shown in bold.}
{\begin{tabular}{S[table-format=2.0, round-precision=0]
|S[table-format=1.3, round-precision=3]S[table-format=1.4, round-precision=4]
|S[table-format=1.3, round-precision=3]S[table-format=1.4, round-precision=4]
|S[table-format=1.3, round-precision=3]S[table-format=1.4, round-precision=4]
|S[table-format=1.3, round-precision=3]S[table-format=1.4, round-precision=4]}
\toprule
{} & \multicolumn{2}{c}{Anticlustering} & \multicolumn{2}{c}{Clustering} & \multicolumn{2}{c}{Random} & \multicolumn{2}{c}{Duplicates} \\
{Arcs} & {$\Delta$D.~obj.} & {$p$-value} & {$\Delta$D.~obj.} & {$p$-value} & {$\Delta$D.~obj.} & {$p$-value} & {$\Delta$D.~obj.} & {$p$-value} \\
\midrule
3 &  &  &  &  &  &  & \bfseries 0.214755 & \bfseries 0.000002 \\
4 & \bfseries 0.023537 & \bfseries 0.036234 & \bfseries 0.079469 & \bfseries 0.000019 & \bfseries 0.030132 & \bfseries 0.001432 & \bfseries 0.293123 & \bfseries 0.000002 \\
5 & \bfseries 0.029991 & \bfseries 0.000036 & \bfseries 0.092195 & \bfseries 0.000002 & \bfseries 0.023555 & \bfseries 0.000006 & \bfseries 0.255869 & \bfseries 0.000002 \\
6 & \bfseries 0.022715 & \bfseries 0.001432 & \bfseries 0.058615 & \bfseries 0.000082 & \bfseries 0.022074 & \bfseries 0.000395 & \bfseries 0.226525 & \bfseries 0.000002 \\
7 & \bfseries 0.017536 & \bfseries 0.000002 & \bfseries 0.031273 & \bfseries 0.000002 & \bfseries 0.017876 & \bfseries 0.000002 & \bfseries 0.202590 & \bfseries 0.000002 \\
8 & \bfseries 0.013701 & \bfseries 0.000004 & \bfseries 0.015578 & \bfseries 0.000006 & \bfseries 0.011805 & \bfseries 0.000002 & \bfseries 0.180329 & \bfseries 0.000002 \\
9 & \bfseries 0.010732 & \bfseries 0.000006 & \bfseries 0.013855 & \bfseries 0.000134 & \bfseries 0.012540 & \bfseries 0.000013 & \bfseries 0.170534 & \bfseries 0.000002 \\
10 & \bfseries 0.009046 & \bfseries 0.000586 & \bfseries 0.009258 & \bfseries 0.000708 & \bfseries 0.005359 & \bfseries 0.002712 & \bfseries 0.139541 & \bfseries 0.000002 \\
11 & \bfseries 0.006728 & \bfseries 0.000019 & \bfseries 0.006692 & \bfseries 0.000168 & \bfseries 0.008194 & \bfseries 0.000105 & \bfseries 0.125833 & \bfseries 0.000002 \\
12 & \bfseries 0.006268 & \bfseries 0.000063 & \bfseries 0.004362 & \bfseries 0.004221 & \bfseries 0.004810 & \bfseries 0.000134 & \bfseries 0.112194 & \bfseries 0.000002 \\
13 & \bfseries 0.006219 & \bfseries 0.000322 & \bfseries 0.003160 & \bfseries 0.026642 & \bfseries 0.005365 & \bfseries 0.001209 & \bfseries 0.113402 & \bfseries 0.000002 \\
14 & \bfseries 0.006344 & \bfseries 0.000105 & \bfseries 0.003054 & \bfseries 0.019234 & \bfseries 0.003805 & \bfseries 0.003153 & \bfseries 0.103380 & \bfseries 0.000002 \\
30 &  &  &  &  &  &  & \bfseries 0.083443 & \bfseries 0.000002 \\
\bottomrule
{-} & {-} & {-} & {-} & {-} & {-} & {-} & {-} & {-}
\end{tabular}
}
{\label{tab:stats_objective_value}}

\statstable
{Median pairwise differences in Gradient Index between the baseline algorithms and the proposed algorithm. Statistically significant results are shown in bold.}
{\begin{tabular}{S[table-format=2.0, round-precision=0]
|S[table-format=1.2, round-precision=2]S[table-format=1.4, round-precision=4]
|S[table-format=1.2, round-precision=2]S[table-format=1.4, round-precision=4]
|S[table-format=1.2, round-precision=2]S[table-format=1.4, round-precision=4]
|S[table-format=1.2, round-precision=2]S[table-format=1.4, round-precision=4]}
\toprule
{} & \multicolumn{2}{c}{Anticlustering} & \multicolumn{2}{c}{Clustering} & \multicolumn{2}{c}{Random} & \multicolumn{2}{c}{Duplicates} \\
{Arcs} & {$\Delta$GI} & {$p$-value} & {$\Delta$GI} & {$p$-value} & {$\Delta$GI} & {$p$-value} & {$\Delta$GI} & {$p$-value} \\
\midrule
3 &  &  &  &  &  &  & \bfseries 1.484056 & \bfseries 0.000006 \\
4 & \bfseries 0.401521 & \bfseries 0.000261 & \bfseries 0.483692 & \bfseries 0.000010 & \bfseries 0.345633 & \bfseries 0.000036 & \bfseries 2.171101 & \bfseries 0.000002 \\
5 & \bfseries 0.411358 & \bfseries 0.000002 & \bfseries 0.556451 & \bfseries 0.000002 & \bfseries 0.335036 & \bfseries 0.000002 & \bfseries 2.024812 & \bfseries 0.000002 \\
6 & \bfseries 0.289968 & \bfseries 0.000002 & \bfseries 0.422163 & \bfseries 0.000002 & \bfseries 0.242570 & \bfseries 0.000002 & \bfseries 1.840201 & \bfseries 0.000002 \\
7 & \bfseries 0.245139 & \bfseries 0.000002 & \bfseries 0.279992 & \bfseries 0.000002 & \bfseries 0.181682 & \bfseries 0.000002 & \bfseries 1.695921 & \bfseries 0.000002 \\
8 & \bfseries 0.164278 & \bfseries 0.000002 & \bfseries 0.148423 & \bfseries 0.000004 & \bfseries 0.136364 & \bfseries 0.000002 & \bfseries 1.684908 & \bfseries 0.000002 \\
9 & \bfseries 0.121746 & \bfseries 0.000019 & \bfseries 0.159784 & \bfseries 0.000048 & \bfseries 0.124876 & \bfseries 0.000002 & \bfseries 1.476054 & \bfseries 0.000002 \\
10 & \bfseries 0.088979 & \bfseries 0.000708 & \bfseries 0.094582 & \bfseries 0.000134 & \bfseries 0.073513 & \bfseries 0.000063 & \bfseries 1.306574 & \bfseries 0.000002 \\
11 & \bfseries 0.070444 & \bfseries 0.000013 & \bfseries 0.071298 & \bfseries 0.000082 & \bfseries 0.093641 & \bfseries 0.000019 & \bfseries 1.372456 & \bfseries 0.000002 \\
12 & \bfseries 0.076035 & \bfseries 0.000210 & 0.023326 & 0.082550 & \bfseries 0.069079 & \bfseries 0.000082 & \bfseries 1.200382 & \bfseries 0.000002 \\
13 & \bfseries 0.078103 & \bfseries 0.000134 & 0.002503 & 0.784126 & \bfseries 0.043876 & \bfseries 0.000395 & \bfseries 1.163238 & \bfseries 0.000002 \\
14 & \bfseries 0.078288 & \bfseries 0.000105 & 0.004426 & 0.784126 & \bfseries 0.054460 & \bfseries 0.000322 & \bfseries 1.088148 & \bfseries 0.000002 \\
30 &  &  &  &  &  &  & \bfseries 1.031547 & \bfseries 0.000002 \\
\bottomrule
{-} & {-} & {-} & {-} & {-} & {-} & {-} & {-} & {-}
\end{tabular}
}
{\label{tab:stats_gradient_index}}

\statstable
{Median pairwise differences in Global Efficiency Index between the baseline algorithms and the proposed algorithm. Statistically significant results are shown in bold.}
{\begin{tabular}{S[table-format=2.0, round-precision=0]
|S[table-format=-1.3, round-precision=3]S[table-format=1.4, round-precision=4]
|S[table-format=-1.3, round-precision=3]S[table-format=1.4, round-precision=4]
|S[table-format=-1.3, round-precision=3]S[table-format=1.4, round-precision=4]
|S[table-format=-1.3, round-precision=3]S[table-format=1.4, round-precision=4]}
\toprule
{} & \multicolumn{2}{c}{Anticlustering} & \multicolumn{2}{c}{Clustering} & \multicolumn{2}{c}{Random} & \multicolumn{2}{c}{Duplicates} \\
{Arcs} & {$\Delta$G$\eta$} & {$p$-value} & {$\Delta$G$\eta$} & {$p$-value} & {$\Delta$G$\eta$} & {$p$-value} & {$\Delta$G$\eta$} & {$p$-value} \\
\midrule
3 &  &  &  &  &  &  & \bfseries -0.046859 & \bfseries 0.000004 \\
4 & \bfseries -0.015224 & \bfseries 0.001209 & \bfseries -0.021492 & \bfseries 0.000002 & \bfseries -0.014337 & \bfseries 0.000134 & \bfseries -0.085947 & \bfseries 0.000002 \\
5 & \bfseries -0.021717 & \bfseries 0.000002 & \bfseries -0.033485 & \bfseries 0.000002 & \bfseries -0.021375 & \bfseries 0.000002 & \bfseries -0.093972 & \bfseries 0.000002 \\
6 & \bfseries -0.014380 & \bfseries 0.000004 & \bfseries -0.026751 & \bfseries 0.000004 & \bfseries -0.016946 & \bfseries 0.000002 & \bfseries -0.090188 & \bfseries 0.000002 \\
7 & \bfseries -0.012213 & \bfseries 0.000002 & \bfseries -0.019268 & \bfseries 0.000004 & \bfseries -0.012102 & \bfseries 0.000002 & \bfseries -0.088969 & \bfseries 0.000002 \\
8 & \bfseries -0.012128 & \bfseries 0.000002 & \bfseries -0.012030 & \bfseries 0.000013 & \bfseries -0.010955 & \bfseries 0.000002 & \bfseries -0.089845 & \bfseries 0.000002 \\
9 & \bfseries -0.009827 & \bfseries 0.000013 & \bfseries -0.010322 & \bfseries 0.000063 & \bfseries -0.009150 & \bfseries 0.000002 & \bfseries -0.085902 & \bfseries 0.000002 \\
10 & \bfseries -0.007512 & \bfseries 0.000082 & \bfseries -0.005837 & \bfseries 0.000322 & \bfseries -0.004450 & \bfseries 0.000048 & \bfseries -0.077606 & \bfseries 0.000002 \\
11 & \bfseries -0.006418 & \bfseries 0.000010 & \bfseries -0.004427 & \bfseries 0.000063 & \bfseries -0.006984 & \bfseries 0.000002 & \bfseries -0.080173 & \bfseries 0.000002 \\
12 & \bfseries -0.007208 & \bfseries 0.000134 & -0.003343 & 0.036234 & \bfseries -0.005544 & \bfseries 0.000210 & \bfseries -0.075417 & \bfseries 0.000002 \\
13 & \bfseries -0.007104 & \bfseries 0.000027 & -0.001555 & 0.142906 & \bfseries -0.005274 & \bfseries 0.001017 & \bfseries -0.071213 & \bfseries 0.000002 \\
14 & \bfseries -0.005813 & \bfseries 0.000210 & -0.000934 & 0.329983 & \bfseries -0.004347 & \bfseries 0.000708 & \bfseries -0.068103 & \bfseries 0.000002 \\
30 &  &  &  &  &  &  & \bfseries -0.066273 & \bfseries 0.000002 \\
\bottomrule
{-} & {-} & {-} & {-} & {-} & {-} & {-} & {-} & {-}
\end{tabular}
}
{\label{tab:stats_global_efficiency_index}}

\statstable
{Median pairwise differences in V$_{12~\unit{\gray}}$ between the baseline algorithms and the proposed algorithm. Statistically significant results are shown in bold.}
{\begin{tabular}{S[table-format=2.0, round-precision=0]
|S[table-format=2.1, round-precision=1]S[table-format=1.4, round-precision=4]
|S[table-format=2.1, round-precision=1]S[table-format=1.4, round-precision=4]
|S[table-format=2.1, round-precision=1]S[table-format=1.4, round-precision=4]
|S[table-format=2.1, round-precision=1]S[table-format=1.4, round-precision=4]}
\toprule
{} & \multicolumn{2}{c}{Anticlustering} & \multicolumn{2}{c}{Clustering} & \multicolumn{2}{c}{Random} & \multicolumn{2}{c}{Duplicates} \\
{Arcs} & {$\Delta$V$_{12~\unit{\gray}}$} & {$p$-value} & {$\Delta$V$_{12~\unit{\gray}}$} & {$p$-value} & {$\Delta$V$_{12~\unit{\gray}}$} & {$p$-value} & {$\Delta$V$_{12~\unit{\gray}}$} & {$p$-value} \\
\midrule
3 &  &  &  &  &  &  & \bfseries 9.288654 & \bfseries 0.000004 \\
4 & \bfseries 2.349904 & \bfseries 0.000395 & \bfseries 3.987677 & \bfseries 0.000002 & \bfseries 2.146549 & \bfseries 0.000134 & \bfseries 15.094992 & \bfseries 0.000002 \\
5 & \bfseries 3.044477 & \bfseries 0.000002 & \bfseries 5.485005 & \bfseries 0.000002 & \bfseries 2.366385 & \bfseries 0.000002 & \bfseries 14.786160 & \bfseries 0.000002 \\
6 & \bfseries 1.760524 & \bfseries 0.000002 & \bfseries 3.088589 & \bfseries 0.000002 & \bfseries 2.092481 & \bfseries 0.000002 & \bfseries 13.541235 & \bfseries 0.000002 \\
7 & \bfseries 1.585177 & \bfseries 0.000002 & \bfseries 2.494187 & \bfseries 0.000004 & \bfseries 1.503848 & \bfseries 0.000002 & \bfseries 13.277099 & \bfseries 0.000002 \\
8 & \bfseries 1.318006 & \bfseries 0.000002 & \bfseries 1.395786 & \bfseries 0.000002 & \bfseries 1.047272 & \bfseries 0.000002 & \bfseries 12.432504 & \bfseries 0.000002 \\
9 & \bfseries 0.990472 & \bfseries 0.000010 & \bfseries 1.418037 & \bfseries 0.000105 & \bfseries 1.066800 & \bfseries 0.000002 & \bfseries 12.209546 & \bfseries 0.000002 \\
10 & \bfseries 0.794161 & \bfseries 0.000134 & \bfseries 0.771844 & \bfseries 0.000483 & \bfseries 0.476543 & \bfseries 0.000027 & \bfseries 10.473444 & \bfseries 0.000002 \\
11 & \bfseries 0.748180 & \bfseries 0.000010 & \bfseries 0.470785 & \bfseries 0.000013 & \bfseries 0.763801 & \bfseries 0.000006 & \bfseries 10.267877 & \bfseries 0.000002 \\
12 & \bfseries 0.597069 & \bfseries 0.000063 & 0.246819 & 0.019234 & \bfseries 0.528435 & \bfseries 0.000168 & \bfseries 9.615847 & \bfseries 0.000002 \\
13 & \bfseries 0.625266 & \bfseries 0.000013 & 0.107038 & 0.113987 & \bfseries 0.488968 & \bfseries 0.000210 & \bfseries 9.555575 & \bfseries 0.000002 \\
14 & \bfseries 0.563824 & \bfseries 0.000027 & 0.224732 & 0.142906 & \bfseries 0.336570 & \bfseries 0.001690 & \bfseries 8.929457 & \bfseries 0.000002 \\
30 &  &  &  &  &  &  & \bfseries 7.462564 & \bfseries 0.000002 \\
\bottomrule
{-} & {\unit{\centi\meter\cubed}} & {-} & {\unit{\centi\meter\cubed}} & {-} & {\unit{\centi\meter\cubed}} & {-} & {\unit{\centi\meter\cubed}} & {-}
\end{tabular}
}
{\label{tab:stats_v12}}

\statstable
{Median pairwise differences in monitor units between the baseline algorithms and the proposed algorithm. Statistically significant results are shown in bold.}
{\begin{tabular}{S[table-format=2.0, round-precision=0]
|S[table-format=-5.0, round-precision=0]S[table-format=1.4, round-precision=4]
|S[table-format=-5.0, round-precision=0]S[table-format=1.4, round-precision=4]
|S[table-format=-5.0, round-precision=0]S[table-format=1.4, round-precision=4]
|S[table-format=-5.0, round-precision=0]S[table-format=1.4, round-precision=4]}
\toprule
{} & \multicolumn{2}{c}{Anticlustering} & \multicolumn{2}{c}{Clustering} & \multicolumn{2}{c}{Random} & \multicolumn{2}{c}{Duplicates} \\
{Arcs} & {$\Delta$MU} & {$p$-value} & {$\Delta$MU} & {$p$-value} & {$\Delta$MU} & {$p$-value} & {$\Delta$MU} & {$p$-value} \\
\midrule
3 &  &  &  &  &  &  & \bfseries 765.794380 & \bfseries 0.000134 \\
4 & 141.476379 & 0.082550 & -130.621067 & 0.545876 & 86.134745 & 0.956329 & -259.048259 & 0.164957 \\
5 & 409.358227 & 0.006390 & \bfseries 744.428587 & \bfseries 0.000168 & \bfseries 597.448635 & \bfseries 0.000483 & \bfseries -1303.191885 & \bfseries 0.000013 \\
6 & \bfseries 650.512470 & \bfseries 0.000210 & \bfseries 773.139182 & \bfseries 0.000027 & \bfseries 546.713935 & \bfseries 0.001209 & \bfseries -1267.797516 & \bfseries 0.000027 \\
7 & 328.986980 & 0.010689 & \bfseries 821.581188 & \bfseries 0.000036 & 308.683203 & 0.023951 & \bfseries -2956.089848 & \bfseries 0.000002 \\
8 & 207.070919 & 0.069580 & \bfseries 620.097693 & \bfseries 0.000036 & 408.792472 & 0.013617 & \bfseries -4430.031781 & \bfseries 0.000002 \\
9 & -7.439981 & 0.474905 & \bfseries 709.365243 & \bfseries 0.000036 & 248.002080 & 0.053169 & \bfseries -6093.207047 & \bfseries 0.000002 \\
10 & -51.159824 & 0.756166 & \bfseries 740.069704 & \bfseries 0.000105 & 187.454064 & 0.113987 & \bfseries -7619.029435 & \bfseries 0.000002 \\
11 & 269.210169 & 0.521673 & \bfseries 802.380798 & \bfseries 0.000004 & 317.414097 & 0.132727 & \bfseries -9011.440397 & \bfseries 0.000002 \\
12 & -194.293235 & 0.202450 & \bfseries 763.026354 & \bfseries 0.019234 & -22.771631 & 0.570597 & \bfseries -10709.540808 & \bfseries 0.000002 \\
13 & 16.747528 & 0.701181 & \bfseries 861.101208 & \bfseries 0.004860 & 445.361484 & 0.277355 & \bfseries -12456.130059 & \bfseries 0.000002 \\
14 & 8.657678 & 1.000000 & \bfseries 938.602645 & \bfseries 0.008308 & 196.789113 & 0.176853 & \bfseries -13754.348012 & \bfseries 0.000002 \\
30 &  &  &  &  &  &  & \bfseries -37204.377819 & \bfseries 0.000002 \\
\bottomrule
{-} & {-} & {-} & {-} & {-} & {-} & {-} & {-} & {-}
\end{tabular}
}
{\label{tab:stats_monitor_units}}

\statstable
{Median pairwise differences in beam-on time between the baseline algorithms and the proposed algorithm. Statistically significant results are shown in bold.}
{\begin{tabular}{S[table-format=2.0, round-precision=0]
|S[table-format=-2.2, round-precision=2]S[table-format=1.4, round-precision=4]
|S[table-format=-2.2, round-precision=2]S[table-format=1.4, round-precision=4]
|S[table-format=-2.2, round-precision=2]S[table-format=1.4, round-precision=4]
|S[table-format=-2.2, round-precision=2]S[table-format=1.4, round-precision=4]}
\toprule
{} & \multicolumn{2}{c}{Anticlustering} & \multicolumn{2}{c}{Clustering} & \multicolumn{2}{c}{Random} & \multicolumn{2}{c}{Duplicates} \\
{Arcs} & {$\Delta$Time} & {$p$-value} & {$\Delta$Time} & {$p$-value} & {$\Delta$Time} & {$p$-value} & {$\Delta$Time} & {$p$-value} \\
\midrule
3 &  &  &  &  &  &  & \bfseries 0.533425 & \bfseries 0.000105 \\
4 & 0.115263 & 0.153646 & -0.102606 & 0.595819 & 0.084797 & 0.869488 & -0.312992 & 0.123093 \\
5 & 0.312972 & 0.007296 & \bfseries 0.516894 & \bfseries 0.000261 & \bfseries 0.465714 & \bfseries 0.000708 & \bfseries -0.814948 & \bfseries 0.000010 \\
6 & \bfseries 0.426828 & \bfseries 0.000261 & \bfseries 0.547112 & \bfseries 0.000013 & \bfseries 0.365948 & \bfseries 0.001432 & \bfseries -0.770398 & \bfseries 0.000027 \\
7 & \bfseries 0.191526 & \bfseries 0.004860 & \bfseries 0.598299 & \bfseries 0.000013 & \bfseries 0.260005 & \bfseries 0.007296 & \bfseries -1.774395 & \bfseries 0.000002 \\
8 & 0.213378 & 0.029575 & \bfseries 0.510623 & \bfseries 0.000010 & \bfseries 0.380257 & \bfseries 0.003153 & \bfseries -2.554015 & \bfseries 0.000002 \\
9 & 0.095911 & 0.058258 & \bfseries 0.592936 & \bfseries 0.000006 & \bfseries 0.347655 & \bfseries 0.000708 & \bfseries -3.331204 & \bfseries 0.000002 \\
10 & 0.006422 & 0.430433 & \bfseries 0.682352 & \bfseries 0.000006 & \bfseries 0.239425 & \bfseries 0.004860 & \bfseries -4.177012 & \bfseries 0.000002 \\
11 & 0.131882 & 0.498009 & \bfseries 0.705740 & \bfseries 0.000002 & 0.288479 & 0.053169 & \bfseries -4.813704 & \bfseries 0.000002 \\
12 & -0.092175 & 0.430433 & \bfseries 0.461483 & \bfseries 0.006390 & -0.015325 & 0.898317 & \bfseries -5.636271 & \bfseries 0.000002 \\
13 & 0.100352 & 0.956329 & \bfseries 0.649984 & \bfseries 0.000708 & 0.456831 & 0.097307 & \bfseries -6.515264 & \bfseries 0.000002 \\
14 & 0.051197 & 0.621513 & \bfseries 0.662302 & \bfseries 0.004221 & 0.240793 & 0.063723 & \bfseries -7.036497 & \bfseries 0.000002 \\
30 &  &  &  &  &  &  & \bfseries -16.730788 & \bfseries 0.000002 \\
\bottomrule
{-} & {\unit{\minute}} & {-} & {\unit{\minute}} & {-} & {\unit{\minute}} & {-} & {\unit{\minute}} & {-}
\end{tabular}
}
{\label{tab:stats_beamon_time}}

\clearpage
% \FloatBarrier
% \supplementarysubsection{Clinical cases}
\subsection{Clinical cases}
\setlength{\intextsep}{0pt}
The six clinical cases are summarized in Table~\ref{tab:case_study_ptvs}.

\statstable
{A summary of the six clinical cases. $\Delta_x$, $\Delta_y$, $\Delta_z$ are the coordinates of the PTV centers relative the case-specific isocenter, and $\left|\Delta_{xyz}\right|$ the distance to isocenter.}
{
\linespread{0.9}\selectfont
\begin{tabular}{ccc|SSSSS}
\toprule
{Case} & {Prescription} & {PTV} & {Volume} & {$\Delta_y$} & {$\Delta_x$} & {$\Delta_z$} & {$\left|\Delta_{xyz}\right|$} \\
\midrule
\multirow[t]{5}{*}{Case 1} & \multirow[t]{5}{*}{3x9 Gy} & PTV 1 & 12.917871 & -4.369267 & -3.989870 & 0.002982 & 5.916888 \\
 &  & PTV 2 & 0.392286 & 3.582503 & -2.828092 & -4.487682 & 6.400916 \\
 &  & PTV 3 & 0.363802 & 0.871202 & 2.916367 & -4.007410 & 5.032248 \\
 &  & PTV 4 & 0.286569 & 1.511525 & 1.267999 & 3.796465 & 4.278513 \\
 &  & PTV 5 & 0.278283 & -1.595964 & 2.633595 & 4.695646 & 5.615337 \\
\midrule
\multirow[t]{6}{*}{Case 2} & \multirow[t]{6}{*}{1x21 Gy} & PTV 1 & 3.134341 & 1.729850 & 4.522333 & -3.957524 & 6.253469 \\
 &  & PTV 2 & 1.536603 & 2.478208 & -2.475938 & 0.677070 & 3.567942 \\
 &  & PTV 3 & 0.493225 & -5.179753 & 4.771296 & -0.148490 & 7.043944 \\
 &  & PTV 4 & 0.478099 & 1.572535 & -6.080697 & 1.510474 & 6.459820 \\
 &  & PTV 5 & 0.476851 & 2.429748 & -0.710406 & 2.332459 & 3.442197 \\
 &  & PTV 6 & 0.270006 & -3.030590 & -0.026588 & -0.413989 & 3.058851 \\
\midrule
\multirow[t]{6}{*}{Case 3} & \multirow[t]{6}{*}{1x18 Gy} & PTV 1 & 0.628476 & 0.810578 & 0.873295 & -1.900624 & 2.243224 \\
 &  & PTV 2 & 0.435408 & -5.586940 & -4.201255 & -0.779644 & 7.033653 \\
 &  & PTV 3 & 0.435332 & 0.973329 & 1.137532 & -0.742879 & 1.671292 \\
 &  & PTV 4 & 0.397699 & 1.034582 & -0.407414 & 0.902726 & 1.432222 \\
 &  & PTV 5 & 0.281700 & 2.866454 & 0.811454 & -2.215496 & 3.712605 \\
 &  & PTV 6 & 0.274540 & -0.098003 & 1.786388 & 4.735917 & 5.062578 \\
\midrule
\multirow[t]{6}{*}{Case 4} & \multirow[t]{6}{*}{5x6 Gy} & PTV 1 & 10.744912 & 2.754328 & -0.779286 & 0.177880 & 2.867970 \\
 &  & PTV 2 & 9.811548 & -3.976259 & 2.041921 & 2.878201 & 5.316400 \\
 &  & PTV 3 & 1.615138 & -1.000476 & -5.283756 & 1.900829 & 5.703699 \\
 &  & PTV 4 & 1.325399 & 2.734807 & 4.459588 & -4.340739 & 6.797728 \\
 &  & PTV 5 & 1.112903 & 1.180363 & 0.947365 & -3.583391 & 3.889917 \\
 &  & PTV 6 & 0.691090 & -1.692764 & -1.385831 & 2.967220 & 3.686512 \\
\midrule
\multirow[t]{9}{*}{Case 5} & \multirow[t]{9}{*}{3x8 Gy} & PTV 1 & 3.415745 & -2.488318 & -1.252660 & -0.435882 & 2.819730 \\
 &  & PTV 2 & 1.673532 & 0.155380 & -2.610387 & 1.117925 & 2.843945 \\
 &  & PTV 3 & 1.290857 & -1.899994 & 4.443268 & 0.491096 & 4.857343 \\
 &  & PTV 4 & 1.018850 & -0.494695 & -4.476916 & -0.279664 & 4.512838 \\
 &  & PTV 5 & 0.705767 & -0.119675 & 3.676843 & 1.767997 & 4.081582 \\
 &  & PTV 6 & 0.476552 & 1.362369 & -1.731784 & -3.567726 & 4.193304 \\
 &  & PTV 7 & 0.416203 & -1.592221 & 0.263237 & 0.974099 & 1.885027 \\
 &  & PTV 8 & 0.366711 & -2.350927 & 1.615257 & 0.278632 & 2.865929 \\
 &  & PTV 9 & 0.196507 & 7.428081 & 0.073143 & -0.346477 & 7.436517 \\
\midrule
\multirow[t]{10}{*}{Case 6} & \multirow[t]{10}{*}{1x21 Gy} & PTV 1 & 2.167978 & -2.119242 & 1.584078 & 0.906562 & 2.796846 \\
 &  & PTV 2 & 1.865381 & 3.312013 & -2.819178 & 2.601411 & 5.067991 \\
 &  & PTV 3 & 1.535213 & 1.586867 & -0.507227 & 3.606332 & 3.972537 \\
 &  & PTV 4 & 0.539755 & -4.936332 & 3.870062 & -3.650007 & 7.257224 \\
 &  & PTV 5 & 0.521938 & -0.334852 & 3.095541 & 0.302619 & 3.128271 \\
 &  & PTV 6 & 0.514059 & -3.704048 & -5.147294 & -1.754804 & 6.579814 \\
 &  & PTV 7 & 0.473219 & -2.748578 & 4.742513 & 0.592047 & 5.513313 \\
 &  & PTV 8 & 0.429964 & 0.805745 & -3.015788 & -1.291576 & 3.378220 \\
 &  & PTV 9 & 0.378084 & 5.334831 & -2.514384 & -3.179227 & 6.700002 \\
 &  & PTV 10 & 0.327074 & 2.803596 & 0.711677 & 1.866644 & 3.442527 \\
\bottomrule
{-} & {-} & {-} & {\unit{\centi\meter\cubed}} & {\unit{\centi\meter}} & {\unit{\centi\meter}} & {\unit{\centi\meter}} & {\unit{\centi\meter}}
\end{tabular}
}
{\label{tab:case_study_ptvs}}

\end{customappendicesenvironment}
}{}

\end{document}